\documentclass{article}
\pdfoutput=1

\usepackage{arxiv}
\usepackage[utf8]{inputenc} 	
\usepackage[T1]{fontenc}    	
\usepackage{hyperref}       	
\usepackage{url}            	
\usepackage{booktabs}       	
\usepackage{amsfonts}       	
\usepackage{nicefrac}       	
\usepackage{microtype}      	
\usepackage{graphicx}			
\usepackage{caption,subcaption} 
\usepackage{float} 			    
\usepackage{todonotes}          
\usepackage{multirow}
\usepackage{authblk}            
\usepackage[load-configurations = abbreviations]{siunitx} 

\graphicspath{{"pictures/"}}
\renewcommand{\arraystretch}{1.5}

\vbadness=\maxdimen

\title{Matrix Element Regression with Deep Neural Networks - breaking the CPU barrier}

\author{Florian Bury}
\author{Christophe Delaere}
\affil{Center for Cosmology, Particle Physics and Phenomenology\\
          Université catholique de Louvain\\
          Louvain-la-Neuve, Belgium \\
          \texttt{florian.bury@uclouvain.be christophe.delaere@uclouvain.be}
}

\begin{document}
\preprint{CP3-20-37}
\maketitle

\begin{abstract}
	The Matrix Element Method (MEM) is a powerful method to extract information from measured events at collider experiments. Compared to multivariate techniques built on large sets of experimental data, the MEM does not rely on an examples-based learning phase but directly exploits our knowledge of the physics processes. This comes at a price, both in term of complexity and computing time since the required multi-dimensional integral of a rapidly varying function needs to be evaluated for every event and physics process considered. This can be mitigated by optimizing the integration, as is done in the MoMEMta package, but the computing time remains a concern, and often makes the use of the MEM in full-scale analysis unpractical or impossible. We investigate in this paper the use of a Deep Neural Network (DNN) built by regression of the MEM integral as an ansatz for analysis, especially in the search for new physics.
\end{abstract}


\section{Introduction}

In the near future the LHC will begin its third run of data collection with the aim of doubling the luminosity collected so far during the two previous runs. While Run-1 allowed the discovery of the Higgs boson, no new physics has yet emerged from Run-2 and  chances are faint to get big surprises in the existing dataset, despite the many analyses that are still ongoing. It is becoming obvious that the expected increase in luminosity will not yield a strong enough gain in sensitivity. Tiny signals hidden behind large backgrounds would not benefit much from additional statistics and would call for an improvement in the analysis techniques to benefit fully from the information contained in each event.

Neural networks and boosted decisions trees, along with other less popular machine learning techniques, are used to learn this information directly from simulations or more rarely from data. The main difficulty of the machine learning methods is to make sure that they actually learned physical information and that they are able to generalize that knowledge. Ensuring this requires using control samples or regularization techniques which have become a standard in the field and many other techniques such as pivoting \cite{pivot}, weakly supervised \cite{WeakSupervised,SemiSupervised} and unsupervised \cite{Unsupervised} learnings have been applied in the field of high energy physics over the past years. The physical information however is provided to the algorithm in an indirect way, by means of the training dataset where it is encoded. Although powerful in terms of prediction, its retro-engineering prospects are limited and therefore restrict its interpretability.

On the contrary, the Matrix Element Method (MEM) uses our knowledge of the Standard Model (SM) by means of the Lagrangian to compute the compatibility of experimental events with a hypothetical process. The underlying knowledge of the physics process and detector response makes its internal dynamic more straightforward to interpret than for machine learning techniques. Since there is no training step, the MEM can also be used when the number of events in the dataset in consideration is very small, a situation in which other methods struggle.

The MEM originates from the Tevatron experiments DØ and CDF for top quark measurement in $t\bar{t}$ production~\cite{MEM1,MEM2,MEM3,MEM4,MEM5,MEM6,MEM7} and has become common in particle physics analyses. Recent examples at the LHC are the searches and measurements of the $t\bar{t}H$~\cite{MEM8,MEM9,MEM10,MEM12,MEM13,MEM14,MEM15} and single top ~\cite{MEM16} production processes, as well as studies of spin correlation in $t\bar{t}$ production~\cite{MEM17}. However this technique suffers from complexity and an expensive computation time. In order to obtain the probability for a given process, the matrix element must be convolved with the parton distribution functions, the transfer function of the detector and be integrated over the whole phase-space. This integral is high dimensional and its integrand has a nontrivial shape with many sharp peaks, the details will be presented in the next section. Even modern implementations of the method (e.g. MoMEMta~\cite{MoMEMta}) that use efficient integration strategies need more CPU time to evaluate this integral than is required by many applications of the method.

The aim of this paper is to show how a Deep Neural Network (DNN) can be used to approximate the result of the matrix element integration. This makes it possible to use the MEM for search for new physics, parameter scans, etc. The probability provided by MoMEMta can be seen as an untraceable function --- meaning there is no closed or computationally affordable form --- of the final state 4-momenta. Any function with reasonable assumptions can be approximated by a neural network given a large enough width --- which can be broken down into several layers ---  and although nothing guarantees the result of the MEM fulfills all these assumptions it was enough to motivate this study. 

\begin{figure}[htp]
	\centering
	\includegraphics[width=.6\textwidth]{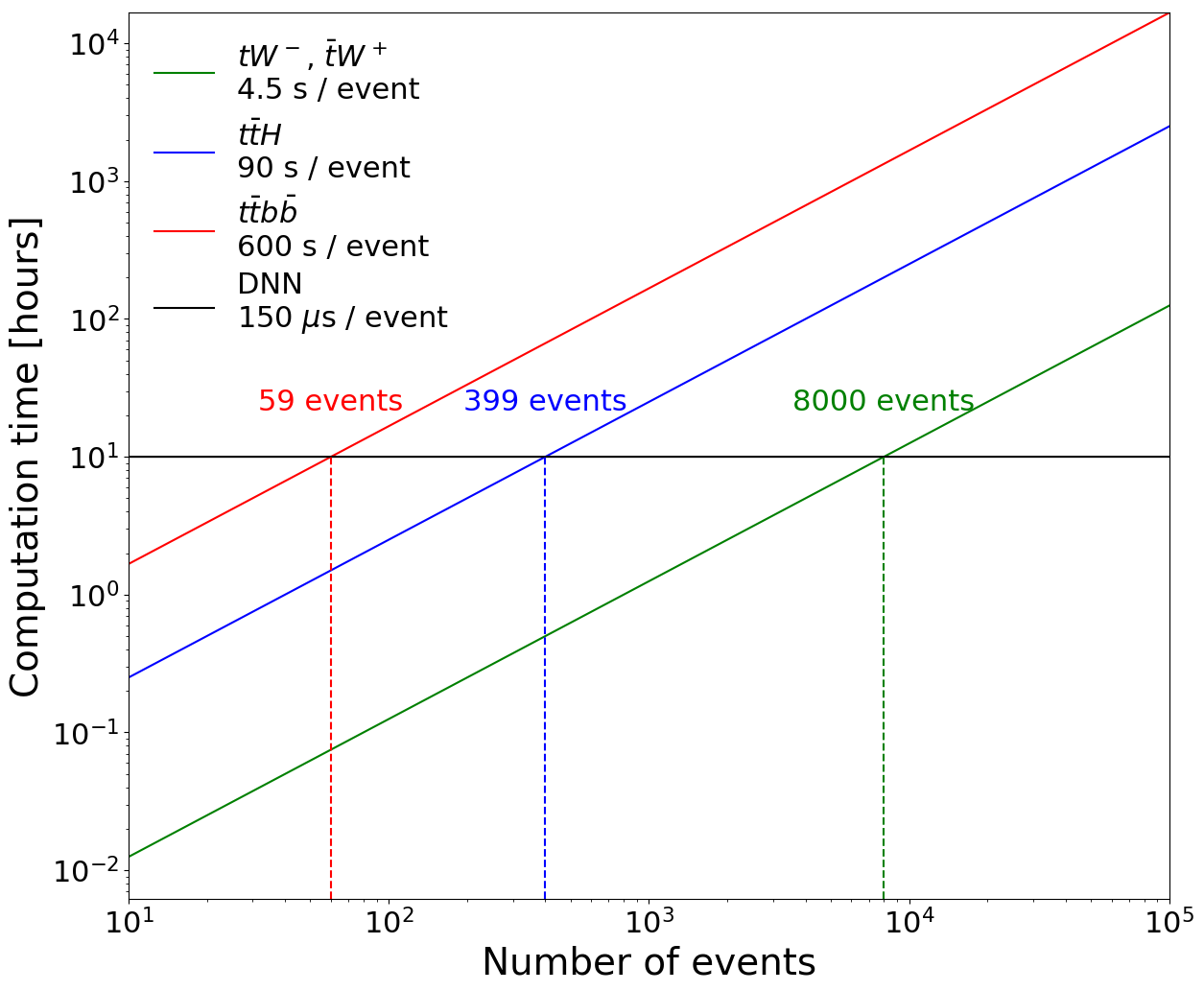}\hfill
	\caption{Computation time as a function of the number of events for few processes with MoMEMta and using the proposed DNN approach. 
	For the DNN it was assumed that the training and evaluation times were \SI{10}{\hour} and \SI{150}{\micro\second} per event respectively.
	Time spent on producing the training sample is not taken into account.
	}
	\label{fig:time_extrapolation}
\end{figure}

This method is potentially much faster that the straight evaluation of the matrix element by integration. A DNN needs a simulated sample and several hours of computing time for training, but evaluating it afterwards is much faster: typically at least a few order of magnitude less computing time than for the classical MEM integration, which takes a few seconds per events even for the easiest processes. This is illustrated in Figure~\ref{fig:time_extrapolation} for a few of the benchmark processes from Ref.~\cite{MoMEMta}.

Many developments have been made recently in the context of the MEM by using either parallel computing \cite{MEMParallel_1,MEMParallel_2} or GPU acceleration \cite{MEMGPU_1,MEMGPU_2}. In addition methods that bypass the classic numerical integration libraries by using boosted decision trees \cite{BDTIntegration} or neural networks \cite{DNNIntegration}, and the recent new applications of normalizing flows for phase-space sampling \cite{NormFlow1,NormFlow2} are promising ways to improve the computation time and potentially to avoid the currently required integration variables optimizations such as implemented in MoMEMta. Nonetheless these techniques can still be coupled with the method we describe in this paper. On the contrary, likelihood-free inference methods \cite{LikelihoodFree_1,LikelihoodFree_2,LikelihoodFree_3} propose to circumvent the integration shortcomings using machine learning to produce the likelihood ratio without any loss of information. Even though apparently more powerful than the MEM, they are less stable and more complicated to interpret than what we propose here.

\section{The Matrix Element Method in a nutshell}
\label{sec:MEM}
The purpose of the MEM is to compute $P(x|\alpha)$, the probability to observe an experimental event given the theoretical hypothesis $\alpha$. In this context $x$ refers to the 4-momenta of an arbitrary number of particles observed in the final state. We will distinguish these experimentally observed particles $x$ from the parton level particles $y$ produced at the interaction point before the hadronization and their detection. $\alpha$ can refer to any set of parameters (e.g. the mass of a resonance) or to different models.

For hadron colliders, the likelihood of a hard scattering producing a partonic final state $y$ is proportional to the differential cross section defined as
\begin{equation}
	d\sigma(q_1,q_2,y) = \frac{(2\pi)^4 |\mathcal{M}(q_1,q_2,y)|^2}{q_1 q_2 s} d\Phi(y),
\end{equation}
where $q_1$ and $q_2$ are the initial state parton momentum fractions and $s$ is the center-of-mass energy. $d\Phi(y)$ is the n-body phase space of the final state $y$, while $|\mathcal{M}(q_1,q_2,y)|^2$ denotes the matrix element for the given process $\alpha$ (including the summation over spin and colors) usually computed numerically at leading order by packages such as MG5\_aMC@NLO~\cite{MG5}.

The propagation of the parton-level 4-momenta $y$ to the experimentally observed ones $x$ includes the parton distribution functions (PDF) $f_{a}(q)$ (for each parton $q$ of flavor $a$), efficiency $\epsilon(y)$ to reconstruct and select the hadronic state $y$ and transfer function $T(x|y)$ normalized with respect to $x$. The latter parameterizes the parton shower, the hadronization and the detector response (whose resolution is limited and produces a smearing of the observed particles momentas). 

The probability $P(x|\alpha)$ results from a convolution of the differential cross section with the transfer function and a sum over initial states: 
\begin{small}
	\begin{eqnarray}
	P(x|\alpha) = \frac{1}{\sigma_{\alpha}^{vis}} \int_{q_1,q_2}  \sum_{a_1,a_2} \int_{y} d\Phi(y) dq_1 dq_2 f_{a_1}(q_1) f_{a_2}(q_2) \frac{(2\pi)^4 |\mathcal{M}(q_1,q_2,y)|^2}{q_1 q_2 s}  T(x|y) \epsilon(y), 	
	\label{eqn:MEM}
	\end{eqnarray}
\end{small}

where $\sigma_{\alpha}^{vis}$ stands for the visible cross-section and is there to make sure that the probability is normalized. It is often computed \textit{a posteriori} with $\sigma_{\alpha}^{vis} = \sigma_{\alpha} \langle\epsilon\rangle$, where $\sigma_{\alpha} = \int d\sigma_{\alpha}(y)$ is the total cross-section and $\langle\epsilon\rangle$ is the average selection efficiency. In practice the integration and the computation of this factor are separated which is why we will omit it in the following from Equation~\ref{eqn:MEM} and define the MEM weights as $W(x|\alpha) = \sigma_{\alpha}^{vis} \times P(x|\alpha)$. In addition the weights can span several orders of magnitude which is why most of the time we will use the event information defined as $I_{\alpha} = -\log_{10}(P(x|\alpha))$ or $I'_{\alpha} = -\log_{10}(W(x|\alpha))$ which only differs by a constant term.

The transfer function includes various complicated processes and several assumptions are usually made to simplify its integration. The detection and measurement of each particle in the final state is independent, at first order, which allows to factorise the transfer function for the different particles. This argument can be pushed even further by factorizing the different components of the measured 4-momentum. We therefore write the transfer function as 
\begin{eqnarray}
T(x|y) &=& \prod_{i=1}^{n} T_i(x^i|y^i), \nonumber\\
T_i(x^i|y^i) &=& T_i^E(x^i|y^i)T_i^{\eta}(x^i|y^i)T_i^{\phi}(x^i|y^i),
\end{eqnarray}
where the index i refers to the final-state particles. 
In most cases, the resolutions in $\eta$ and $\phi$ are very narrow and are parameterized as delta functions. On the contrary, the energy resolution depends on the nature of the particles and should reproduce the behavior of the  simulation detector effects, typically Gaussian for fast simulations. Note that these assumptions can break down when two objects have small angular separation which would require specific care to not impair the convergence and accuracy of the integration. In this paper we derived custom simulation-based binned transfer functions to account for potential asymmetries. 

The high-dimension integration in Equation~\ref{eqn:MEM} requires the use of numerical integrators such as Vegas~\cite{VEGAS}. These tools implement adaptive Monte Carlo techniques, the basic principle of which is to randomly generate points evaluated with the function that one wishes to integrate. With enough points, a relatively close approximation of the integral can be obtained along with its uncertainty. However this method becomes extremely expensive in high-dimension space: while flat regions of the phase-space only need a few points in order to get a good approximation of their integrals, regions where the function fluctuates a lot need to be well covered. 

Adaptive Monte Carlo techniques are iterative processes designed to populate the phase-space heterogeneously in order to decrease the integral variance. Even though they perform better than the uniform sampling, they do not scale easily with the dimensionality and the factorization assumption on which some are based --- e.g. importance sampling in Vegas --- makes them especially suboptimal if the integrand presents peaks that depend on several integration variables. In our case sharp peaks can arise from the propagators in the matrix element or in narrow  transfer function. The latter can already be mapped to integration variables in the classic parameterization of the phase-space
\begin{equation}
d\Phi = \left( \prod_{i=3}^{n} \frac{|\mathbf{p}_i|^2d|\mathbf{p}_i| sin \theta_i d\theta_i d\phi_i}{2 E_i (2\pi)^3} \right) (2\pi)^4 \delta^4\left(p_1+p_2 -\sum_{j=3}^{n}  p_j  \right),  
\label{eqn:phase_space}
\end{equation}
where the Dirac function ensures the momentum conservation. Propagator enhancements then need to be addressed by inverting the Breit-Wigner resonances and the delta functions need to be integrated out. In addition this parameterization includes invisible particles, i.e. particles that did not leave a trace in the detector. Neutrinos, particles outside the acceptance and initial-state partons should be taken into account and the large volume that they represent will heavily impact the integration - unless the kinematic constrains are used to remove these degrees of freedom.

\section{Fitting the MEM with DNN} 

Computing the weight of an event with MoMEMta can take from a few seconds to several minutes depending on the complexity of the process in question. 
This has to be repeated for each hypothesis $\alpha$ and for each event and quickly becomes prohibitive.
In a real-life analysis with several hypotheses and sometimes additional model parameters, it often makes the implementation of the method challenging.
In a nutshell, the approach proposed in this study uses MoMEMta to process simulated samples and produce event information $I'$ under different hypotheses. The result is then used together with MoMEMta inputs --- the 4-momentas of each visible particle --- to train a DNN. As shown in what follows, the resulting network can be used instead of MoMEMta on a larger set of events and for different values of the model parameters.

The inputs of MoMEMta are the 4-momenta of all the observed particles as well as the missing transverse energy (only its $P_T$ and $\phi$ angle), consequently these are the inputs that we want to provide to the DNN. Some rather standard pre-processing has been used to facilitate its task, in the spirit of Ref.~\cite{DataPreProcess}. Depending on the longitudinal difference in momentum between the initial partons that collide in the detector, the particles produced might be boosted in one or the other direction. The network will still be able to learn the function at Equation~\ref{eqn:MEM} but will also have to learn about the longitudinal boost itself, which hinders its ability to describe the interesting part of the matrix element. Using as inputs the $P_T$, $\eta$ and $\phi$ angles for each of the particles improves a lot the situation, since the $\Delta \eta$ between two particles is in good approximation independent of the longitudinal boost. Furthermore, the detector has a cylindrical symmetry and we do not want the network to learn about an arbitrary reference in $\phi$. Any relative quantity defined on the angles could in principle be used, for example the relative $\Delta \phi$ angle with an arbitrary selected particle. This parameterization has shown to yield better results than the raw 4-momenta without loss of generality. While in the integration of the MEM the ordering of the particles is important because all permutations of indistinguishable particles -- e.g. two jets originating from a quark and an antiquark --- need to be considered, there is no notion of distance in the inputs of a fully connected network and it does not have to be taken into account. As the targets span several orders of magnitude we will follow the example set by Ref.~\cite{MoMEMta} and regress on the event information $ I' = -\log_{10}(\textrm{weight})$ instead. Similar approaches have been studied in detail in the literature \cite{outputPreprocess}.

In some cases, the computation does not converge before reaching the maximum number of iterations. In these cases the weights have non-physical infinitesimal values with even smaller uncertainties, referred to as \textit{invalid}. These weights are logically not included in the learning process, but the DNN can be evaluated on these events as on any other.
In order to probe both the behavior of the MEM and the DNN for these events, some of the invalid weights can be recomputed in MoMEMta with more sampling points and iterations and evaluated with the DNN trained on valid weights.

An inherent quality of the DNN approach is the ability to interpolate on inputs that were not seen during the training. In the classical method, parameter scans require to generate weights at different parameter values and eventualy to perform an interpolation. But the cost grows with the granularity and the dimensionality of the parameter space, which can be prohibitive. The advantage of the DNN is that no integration with MoMEMta is necessary anymore once the model is built, even for new events and new parameter values --- as long as one stays clear of the extrapolation regime where DNN are know to be untrustworthy.

Note that in practice the MEM probabilities --- and by extension the MEM information $I'$ --- are rarely used directly in an analysis and are combined in an application-dependent procedure. A simple comparison of their values is thus not a sufficient criterion to state that the method we propose can be used without losses. 

We have used the Keras~\cite{keras} library interfacing Tensorflow~\cite{tensorflow} to train the DNNs. The datasets were separated into three sets: one for the training ($\sim 70\%$), one for the hyperparameter scans used in model selection ($\sim 10\%$) and one for the performance evaluation of the selected model ($\sim 20\%$). All the plots in the paper contain events in the last set. All this is done to detect any overfitting of the network, i.e. the loss of generalization to unknown data because statistical fluctuations of the training data are learnt in addition to, or instead of, the general features of its underlying distribution.

\section{Proof of concept : the \texorpdfstring{$llb\bar{b}$}{llbb} topology}
\label{sec:2HDM}

As a proof of concept we will apply the method we propose in this paper to several processes producing two opposite sign leptons and two jets initiated by b quarks (bjets) as detected particles. This topology is interesting because the main contributions to this topolgy --- Drell-Yan $Z\rightarrow l^+l^-$ production with additional jets, and top quark pair production $t\bar{t}$ with leptonic decays of the W bosons from both top quarks --- are very dissimilar in the way they are treated in the integration. 
The computation for the former is rather straightforward as no missing particles are produced while the latter contains undetected neutrinos whose degrees of freedom need to be accounted for. 
We will then consider the resonnant $H \rightarrow Z (\rightarrow l^+ l^-) A (\rightarrow b \bar{b})$ process that arises in the context of Two Higgs Doublet Models (2HDM), and has been studied by the ATLAS~\cite{AtlasHZA} and CMS~\cite{HZA} collaborations.
The multiple resonances in this process present an interesting challenge from the integration point of view and the unknown masses of the $H$ and $A$ bosons will illustrate the power of our method when the parameter space is multi-dimensional - which is precisely where the classical integration is impractical.  

A summary of the number of events for which the weights have been computed is given in Table~\ref{tab:size}. Only 500K events for the $t\bar{t}$ process have been used in the training to not unbalance the network. The $H \rightarrow ZA$ samples and weights are split in 23 mass configurations up to \SI{1}{\TeV} for both $m_A$ and $m_H$.
The event information $I'_\alpha$ will be evaluated for every event --- $t\bar{t}$, Drell-Yan and $H \rightarrow ZA$ --- 23 times with different mass parameters for the $H \rightarrow ZA$ case alone.
The number of invalid weights and how many were recomputed with more iterations are also given in the table. The non-convergence indicates that it is very difficult for MoMEMta to associate an event with the process the weights is computed for, likely because the event is too incompatible with the hypothesis. The changes of variables introduced in the computation are an attempt to optimize the axes of the integration to the kinematics of the process in question. If the event is pathological or comes from another process then the integrator might not generate points in the large value regions. Therefore more iterations or more points are needed and the integration can reach the threshold on the number of steps before convergence. Sometimes the threshold might be too hard and the convergence was very close, this is probably the case for the invalid Drell-Yan weights given their relatively small number. In addition the fact that all of them converged at the recomputation step with more time and points tends to support this explanation. On the other hand, it is possible that the phase-space regions to be populated for the event are very far in the tails of the sampling distribution for the process under investigation. This is probably the case for a portion of the $t\bar{t}$ weights that did not converge even with more points, especially for $H \rightarrow ZA$ events with invalid weights that are mostly at high $M_H$ and $M_A$ (two thirds of the them are for cases where both masses are at the \SI{}{\TeV} scale). This is because the kinematics of the leptons and bjets are incompatible with a \SI{173}{\GeV} precursor, and thus make the task of MoMEMta more complicated. This does not happen for the Drell-Yan weights because the kinematic range of the products is more flexible. A study of these invalid weights is presented in a dedicated section.

\begin{table}
	\caption{Sample sizes for the three types of weights and the three samples. \textit{Valid} represents weights that converged, \textit{Invalid} the ones that did not and \textit{Invalid recomputed} the ones in the invalid category that converged when recomputed.}
	\centering
	\resizebox{\textwidth}{!}{
	\begin{tabular}{|l|ccc|ccc|cc|}
		\hline
		Process & \multicolumn{3}{|c|}{Drell-Yan weights} & \multicolumn{3}{|c|}{$t\bar{t}$ weights} & \multicolumn{2}{|c|}{$H \rightarrow ZA$ weights} \\
		\hline
		Weights & Valid & Invalid & Invalid & Valid & Invalid & Invalid & Valid & Invalid \\
		&   &   &  recomputed &  &   &  recomputed &  &  \\
		\hline
		Drell-Yan sample & 305407 & 12 & 12 & 305407 & 913 & 480 & 38642 & 3842 \\
		$t\bar{t}$ sample & 2903472 & 109 & 109 & 2903472 & 751 & 545 & 39441 & 14441 \\
		$H \rightarrow ZA$ sample & 209130 & 10 & 10 & 209130 & 37084 & 10040 & 232719 & 32274 \\
		\hline
	\end{tabular}}
	\label{tab:size}
\end{table}

We made sure the events were weighted proportionally to the sample size between the three categories so that they all have the same importance in the training of the DNN. The goal for the $H \rightarrow ZA$ weights is to include $M_H$ and $M_A$ in the inputs of the DNN and provide the Information $I'$ as target. This parametric DNN~\cite{paramDNN} would be very interesting for parameter scans (for example in a maximum likelihood context) that become prohibitive with the classical approach. This DNN is therefore capable to provide Information for parameter values it had not seen during the training and the interpolation that must be performed apart from the integration in the classical approach is now embedded in our method. 

In the next section we will detail the computation process and regression results of each type of weights. The summary of computation time and DNN topologies are at Table~\ref{tab:DNN_comp}.

\subsection{Drell-Yan weights}
\label{sec:DY_weights}
Drell-Yan weights are the easiest to compute since the topology of the process is quite flexible in the range of allowed kinematics for the products. This makes the task of MoMEMta relatively easy when the correct changes of variables is applied. In practice the evaluation of the Drell-Yan weights takes on average a few seconds per event, some rare events can take a few tens of seconds in the tail of the distribution. 

\begin{figure}[htp]
	\centering
	\includegraphics[width=.33\textwidth]{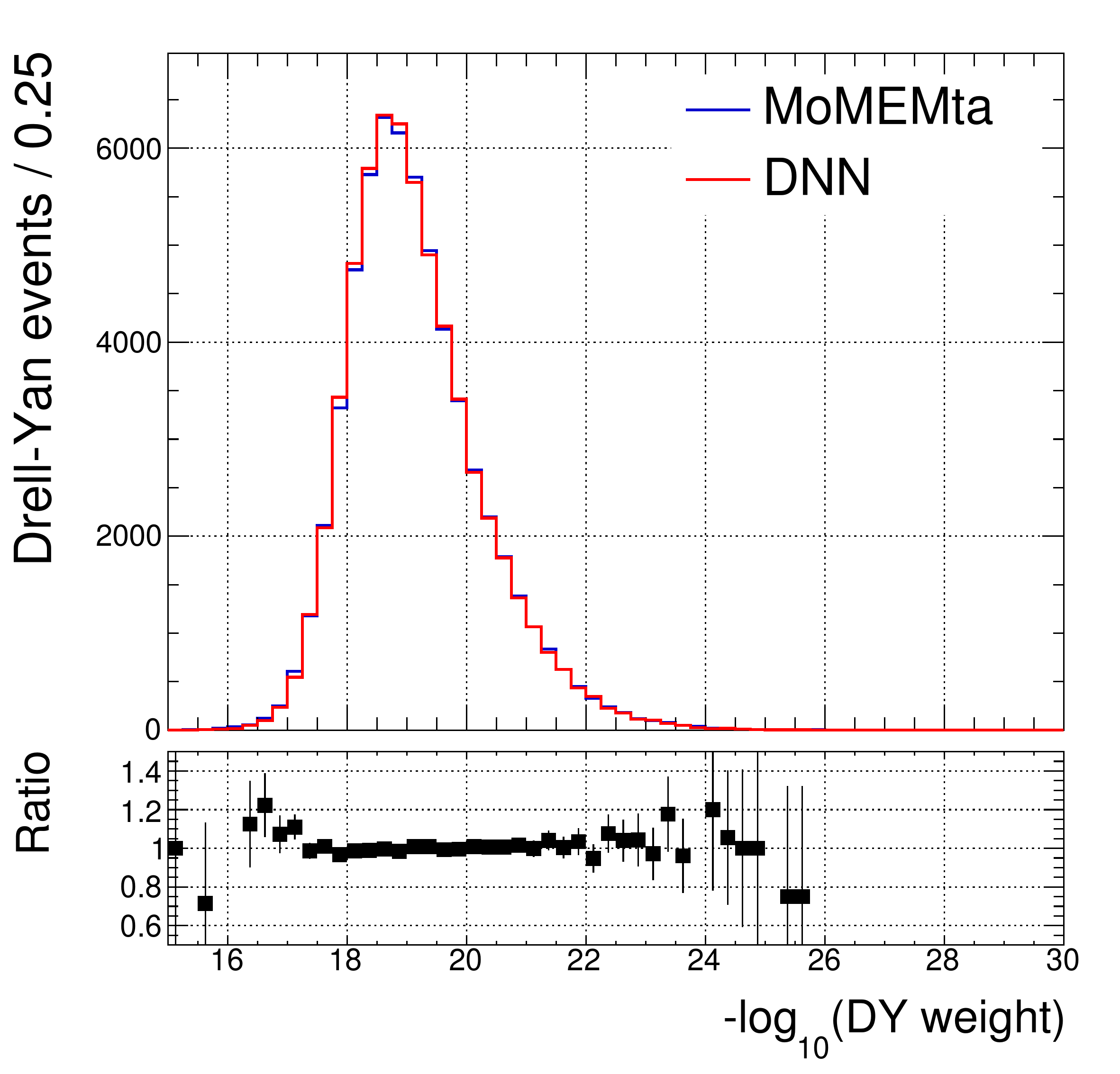}\hfill
	\includegraphics[width=.33\textwidth]{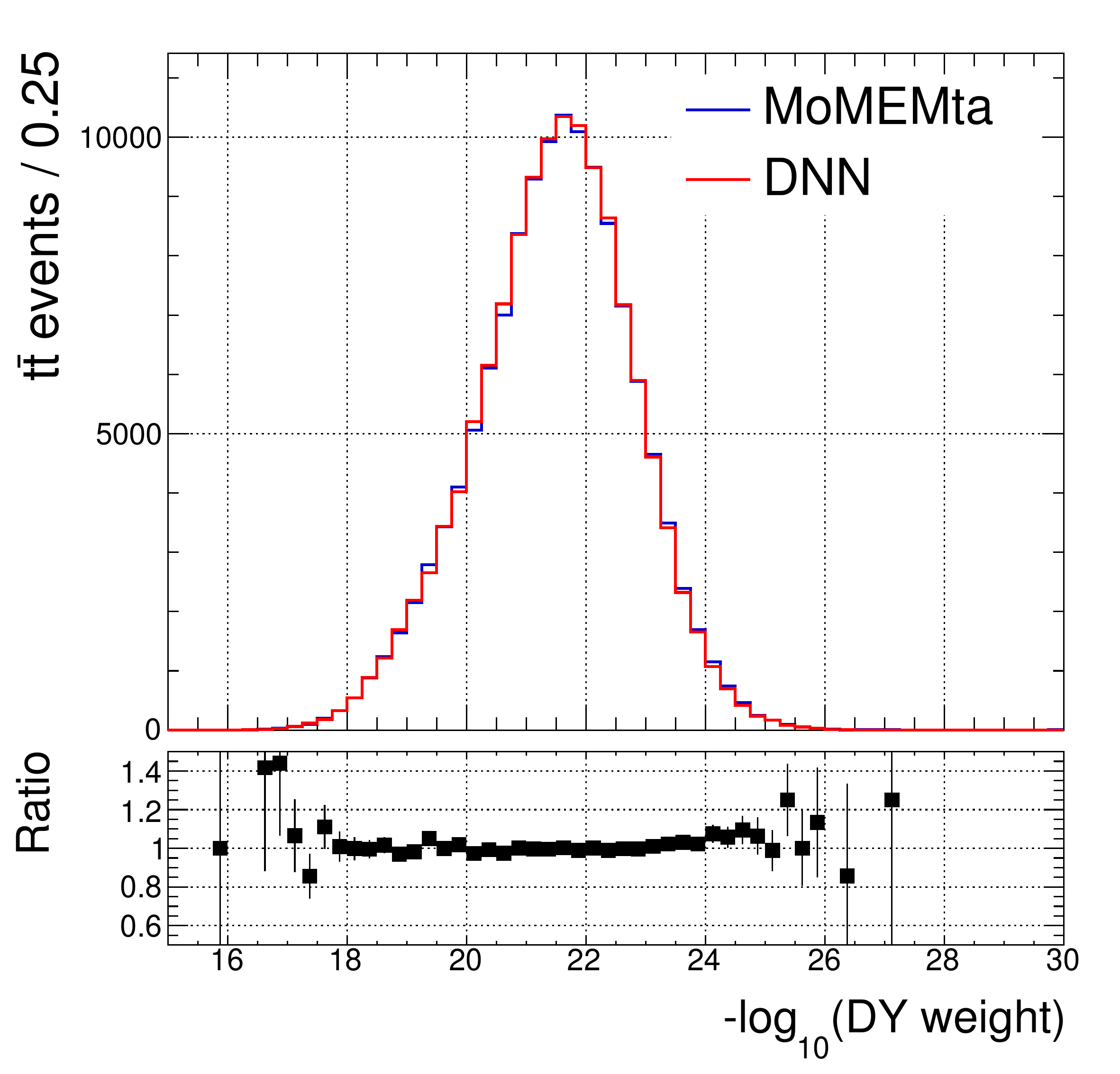}\hfill
	\includegraphics[width=.33\textwidth]{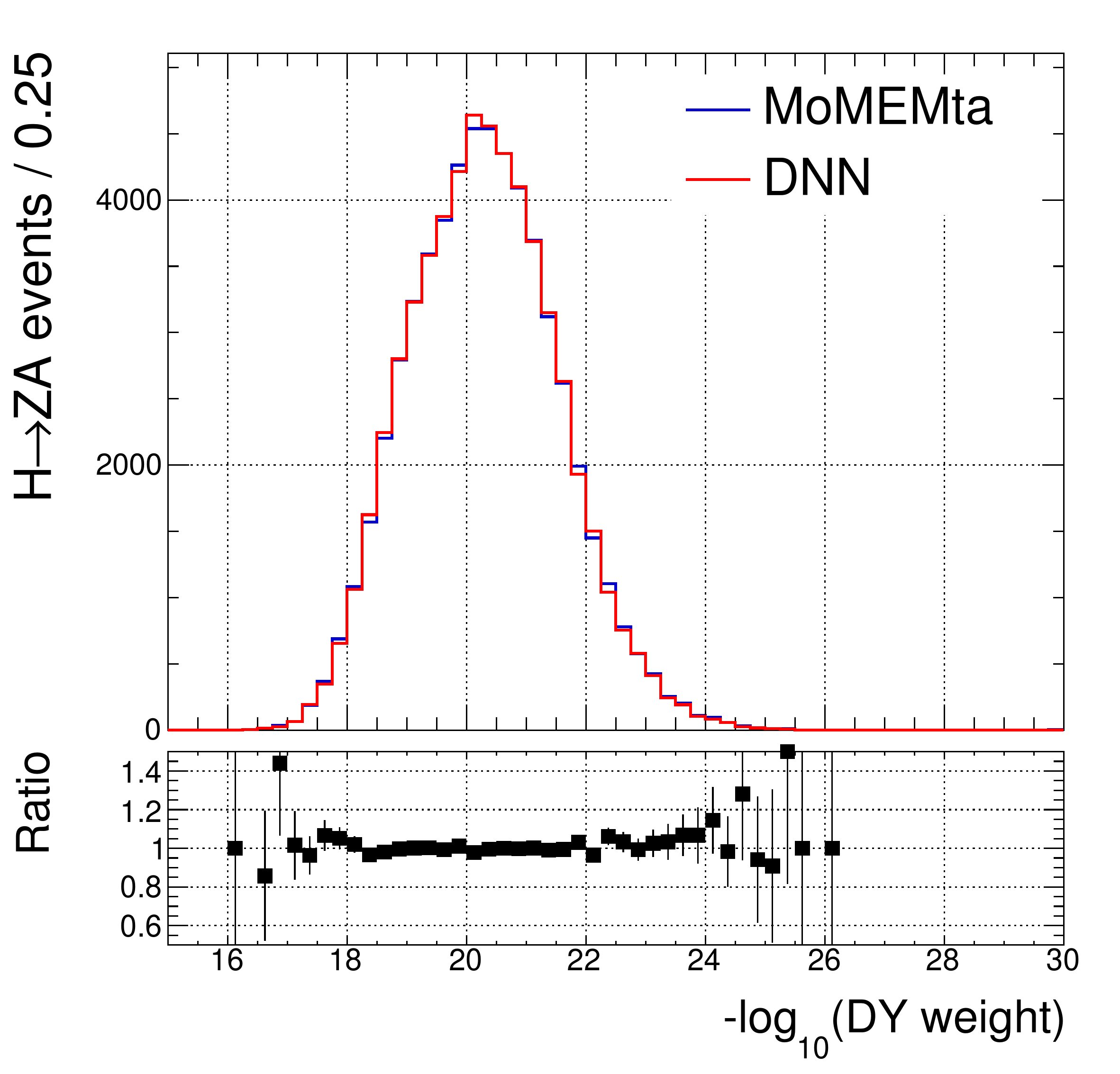}
	\caption{Distributions of the event information $I'_{DY}$ from both MoMEMta and the DNN for the three samples : Drell-Yan (left), $t\bar{t}$ (middle) and $H \rightarrow ZA$ (right) events.}
	\label{fig:DY_weight}
\end{figure}

The $I'_{DY}$ distributions are given in Figure~\ref{fig:DY_weight} for the three types of sample. The agreement is very good between the weights from the MEM computed with MoMEMta and the ones from the DNN. The best model selected during the parameters scan is the one with six layers of 200 neurons, \textit{relu}~\cite{relu} and \textit{selu}~\cite{selu} activation functions for the hidden and output layers respectively, the optimizer for the gradient descent was Adam~\cite{adam}. Dropout~\cite{dropout} and L2~\cite{l2} regularization technique did not improve the efficiency of the training. We emphasize that these events have never been seen by the DNN during the training or in the model evaluation in order to detect overfitting. As expected the Drell-Yan events have higher weights than $t\bar{t}$ and $H \rightarrow ZA$ ones because they have more compatibility with the Drell-Yan hypothesis. 

\subsection{\texorpdfstring{$t\bar{t}$}{ttbar} weights}
\label{sec:TT_weights}
Due to the more complicated topology of the $t\bar{t}$ process --- with fully leptonic decay, this will be implicit from now on --- from the narrow top resonance, the $t\bar{t}$ weights are more intricate to compute. However taking advantage of the Breit-Wigner resonances in a change of variables, the computation time can be reduced to a reasonable level (about 3.2 times slower than for the Drell-Yan weights).

\begin{figure}[htp]
	\centering
	\includegraphics[width=.33\textwidth]{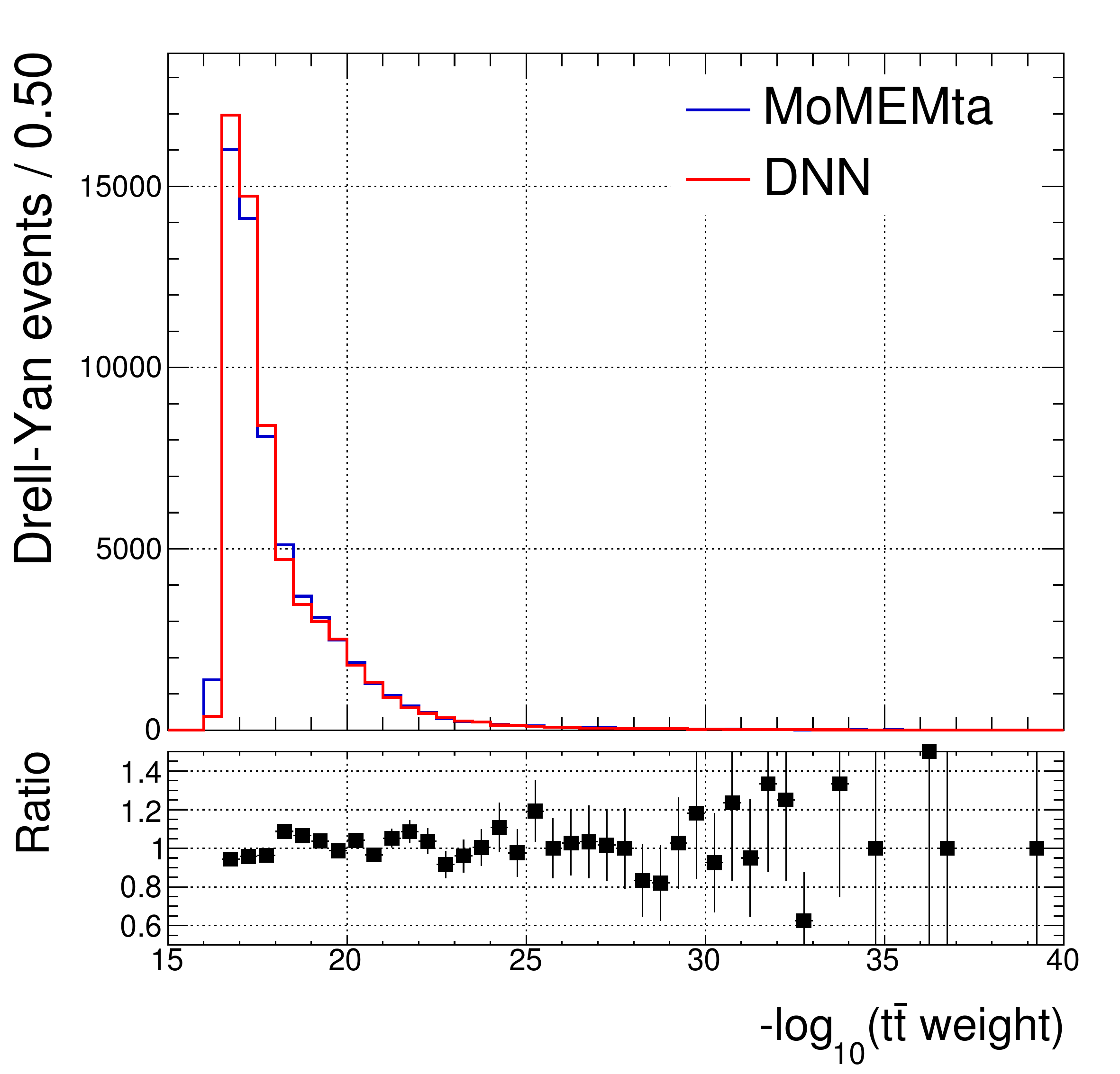}\hfill
	\includegraphics[width=.33\textwidth]{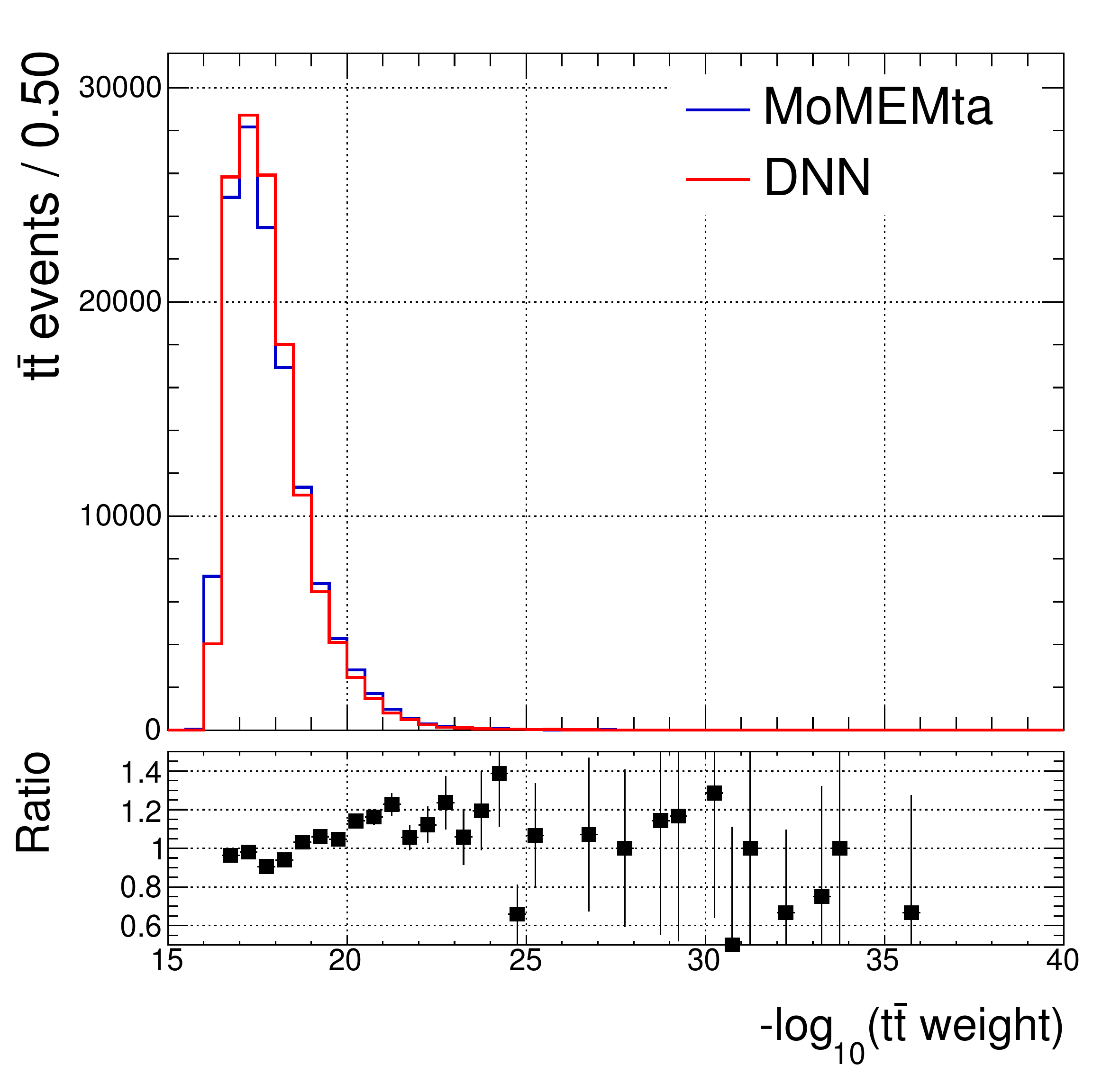}\hfill
	\includegraphics[width=.33\textwidth]{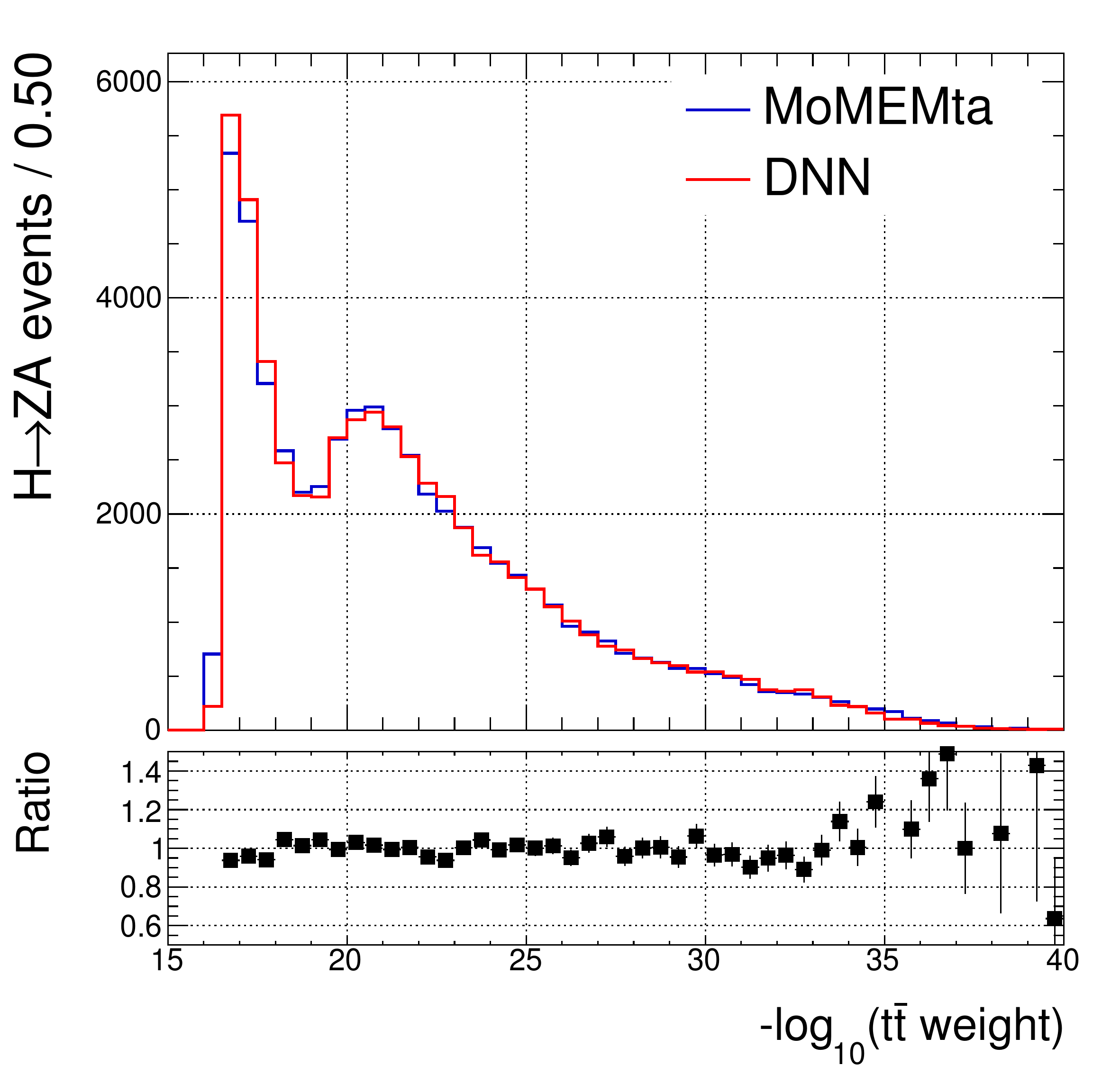}
	\caption{Distributions of the event information $I'_{t\bar{t}}$ from both MoMEMta and the DNN for the three samples : Drell-Yan (left), $t\bar{t}$ (middle) and $H \rightarrow ZA$ (right) events.}
	\label{fig:TT_weight}
\end{figure}

The corresponding $I'_{t\bar{t}}$ distributions are shown in Figure~\ref{fig:TT_weight}. The contrast in the weight distribution between the Drell-Yan and $t\bar{t}$ samples is less obvious but a longer tail can be observed for the Drell-Yan events. The double peak of the $H \rightarrow ZA$ case comes from the different mass configurations $M_H$ and $M_A$ that constitute this sample. High (pseudo)scalar masses lead to low weights while low masses are more consistent with the $t\bar{t}$ hypothesis. Overall the agreement between the classically computed weights and the ones from the DNN is good. The best model is close to the one for the Drell-Yan weights: it contains eight layers of 500 neurons and a small L2 regularization factor, probably to counter the overfitting of such a deep network.

\subsection{\texorpdfstring{$H\rightarrow ZA \rightarrow llbb$}{H->llbb} weights}
\label{sec:signal_weights}

The case of the $H \rightarrow ZA$ hypothesis is more complicated due to its hypothetical nature that makes this process dependent on unknown parameters. In our case we only varied two parameters, the masses $M_H$ and $M_A$, while other have been fixed. We have focused on 23 configurations, both for the event generation and for the MEM computation. While some integration tricks were beneficial for the other hypotheses, mostly by using the delta function for the momentum conservation in Equation~\ref{eqn:MEM} to remove some degrees of freedom --- such as the bjets momentum magnitude for the Drell-Yan process or replace the integration over the invisible particles by integrating over the resonances in the $t\bar{t}$ process --- the absence of invisible particles and the multiple resonances of the $H \rightarrow ZA$ process prevent these tricks to be profitable . This has a heavy impact on the computation time, about 50000 CPU days for all the samples (on average about ten minutes per event per weight).

\begin{figure}[htp]
	\centering
	\includegraphics[width=.5\textwidth]{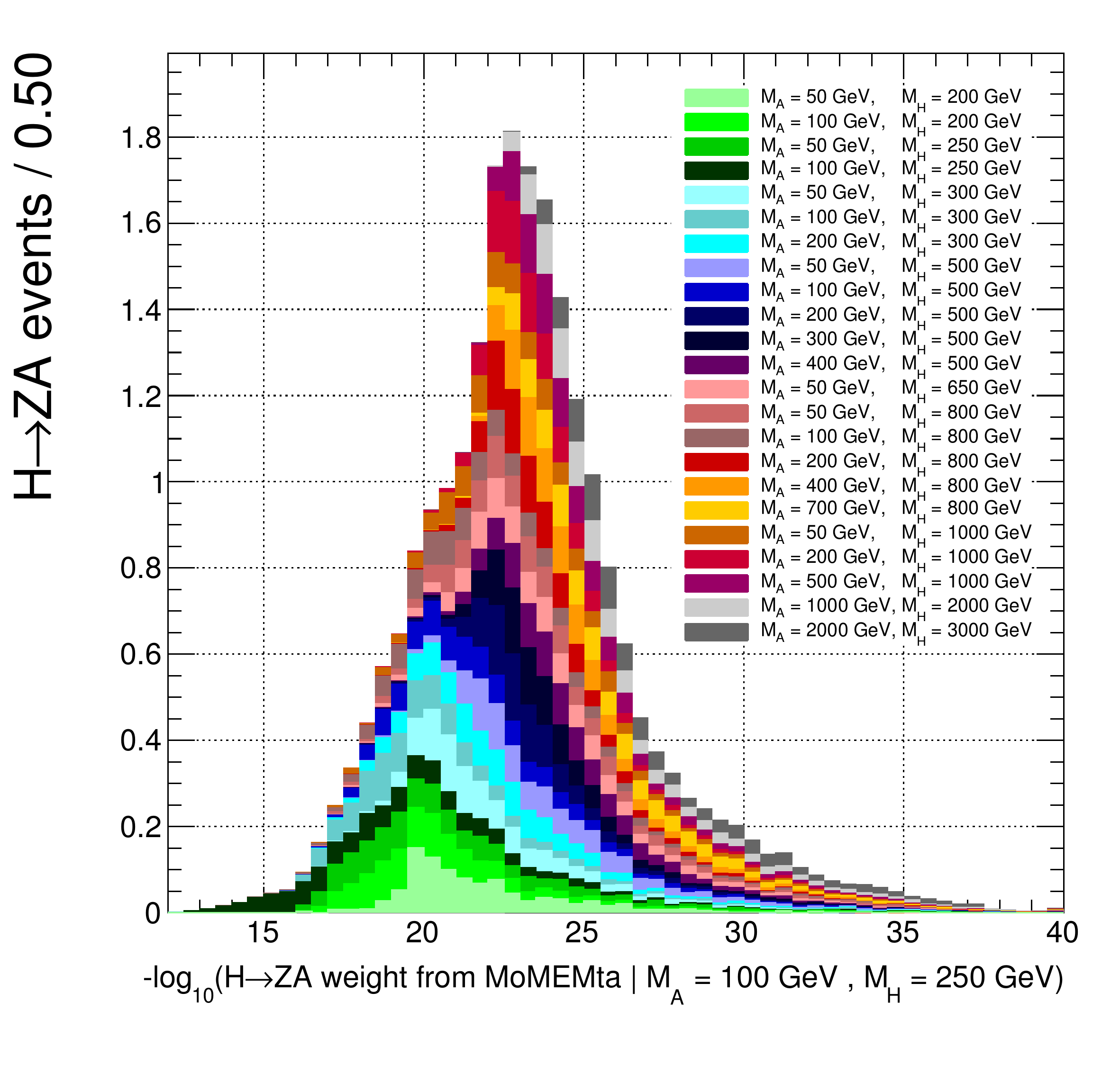}\hfill
	\includegraphics[width=.5\textwidth]{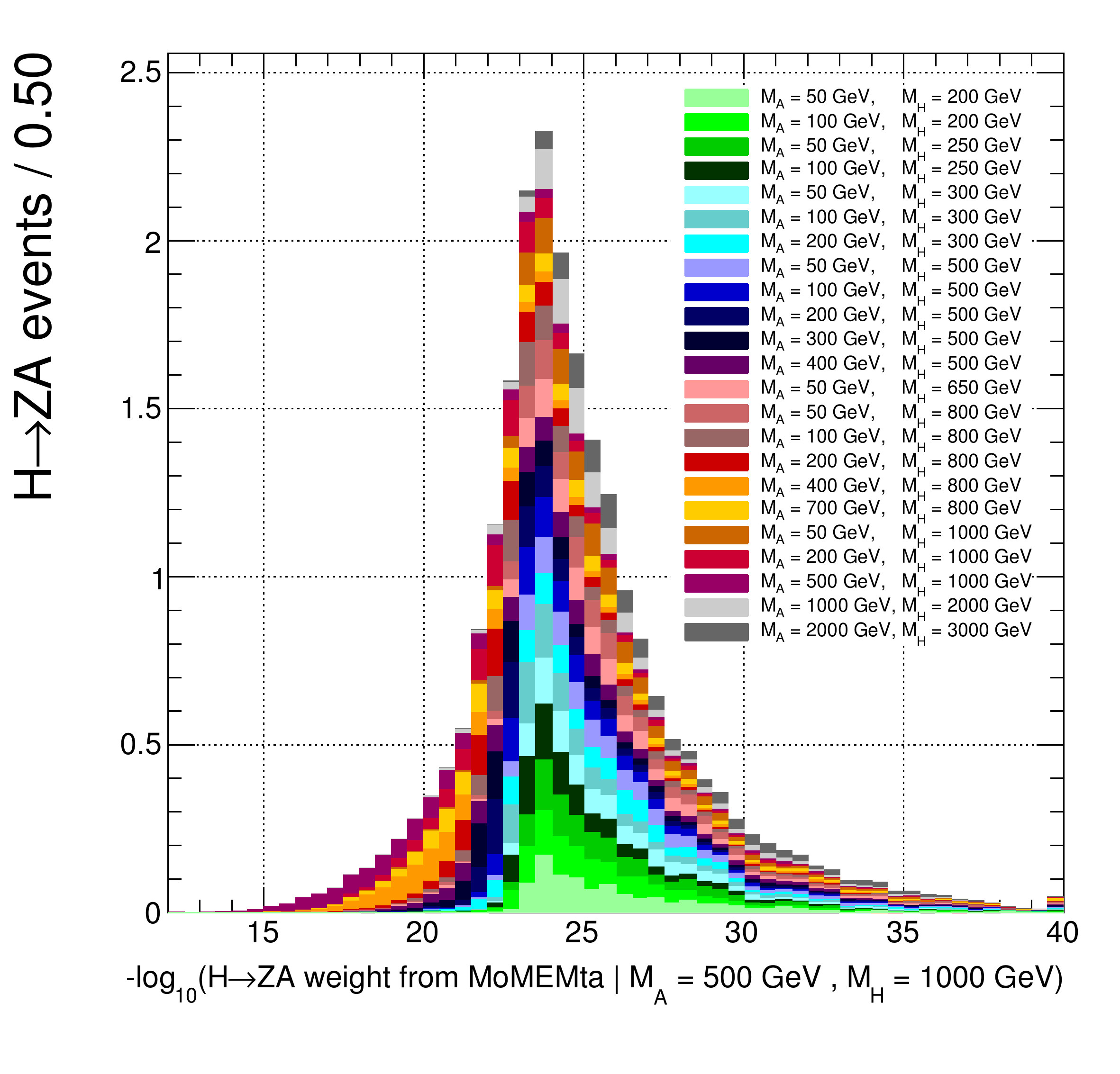}
	\caption{Distributions of the event information $I'_{H\to ZA}$ for two mass hypotheses --- ($M_A = \SI{100}{\GeV}$ , $M_H = \SI{250}{\GeV}$) (left) and ($M_A = \SI{50}{\GeV}$ , $M_H = \SI{1000}{\GeV}$) (right). On each plot, the contributions from the samples with different masses used for the event generation have been stacked on top of each other for clarity.}
	\label{fig:signal_stack}
\end{figure}

Figure~\ref{fig:signal_stack} shows the different contributions of the weight distributions from the event generated with different resonant masses. 
As expected, the event information is lower (hence the probability is higher) when the masses used for the generation match the hypothesis used for the weight calculation, although the dependence on the visible cross-section is not taken into account here and might have a small effect. 
This can be seen on the left part of Figure~\ref{fig:signal_stack} where the low mass events have higher weights. 
On the contrary, the high mass events have the highest weights in the right part of the figure, as they match more closely the topology of the high mass weights. Interestingly enough, events with same $M_A$ but different $M_H$ compared to the values included in the weight tend to have slightly higher probabilities.

\begin{figure}[htp]
	\centering
	\includegraphics[width=.33\textwidth]{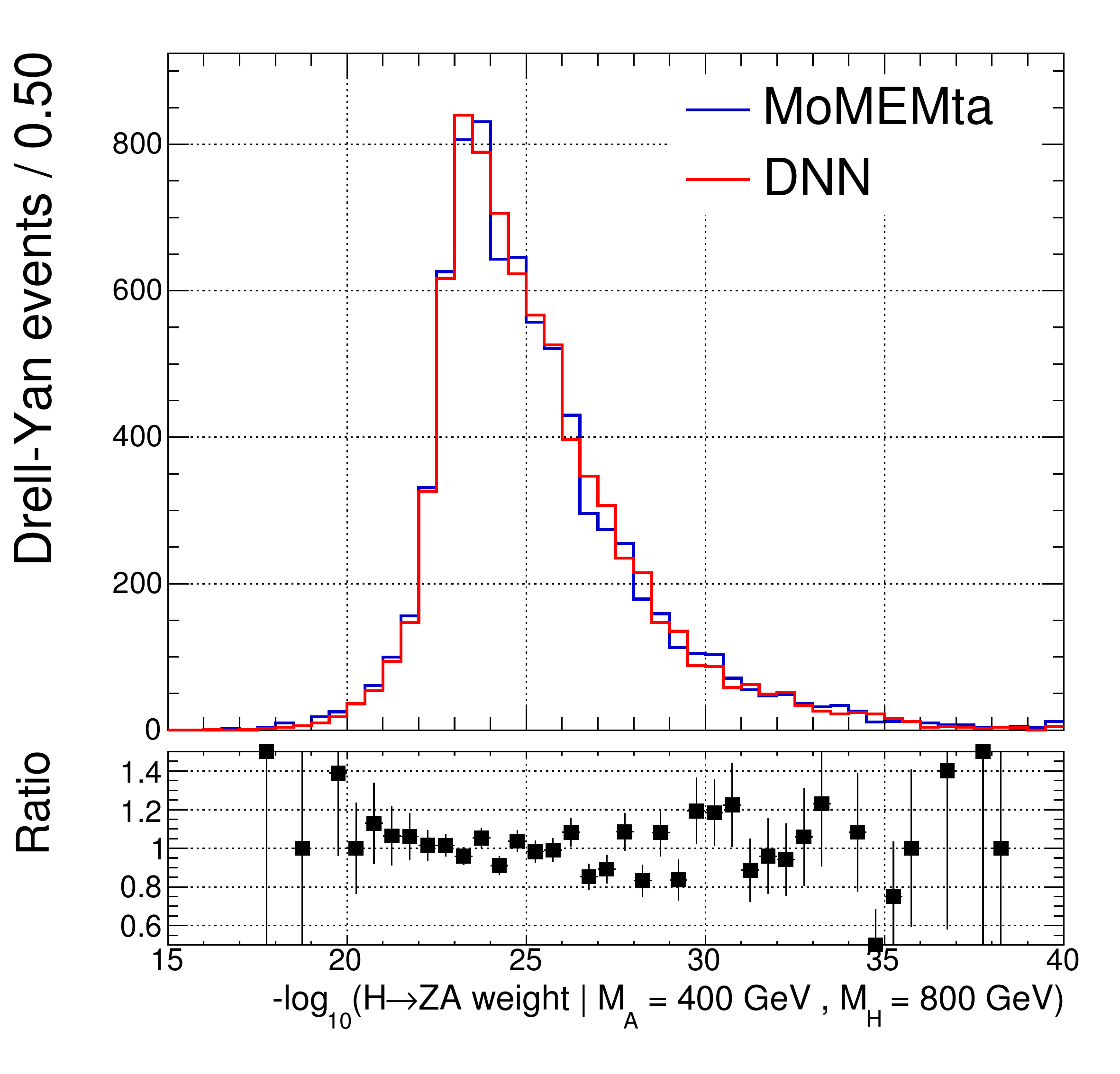}\hfill
	\includegraphics[width=.33\textwidth]{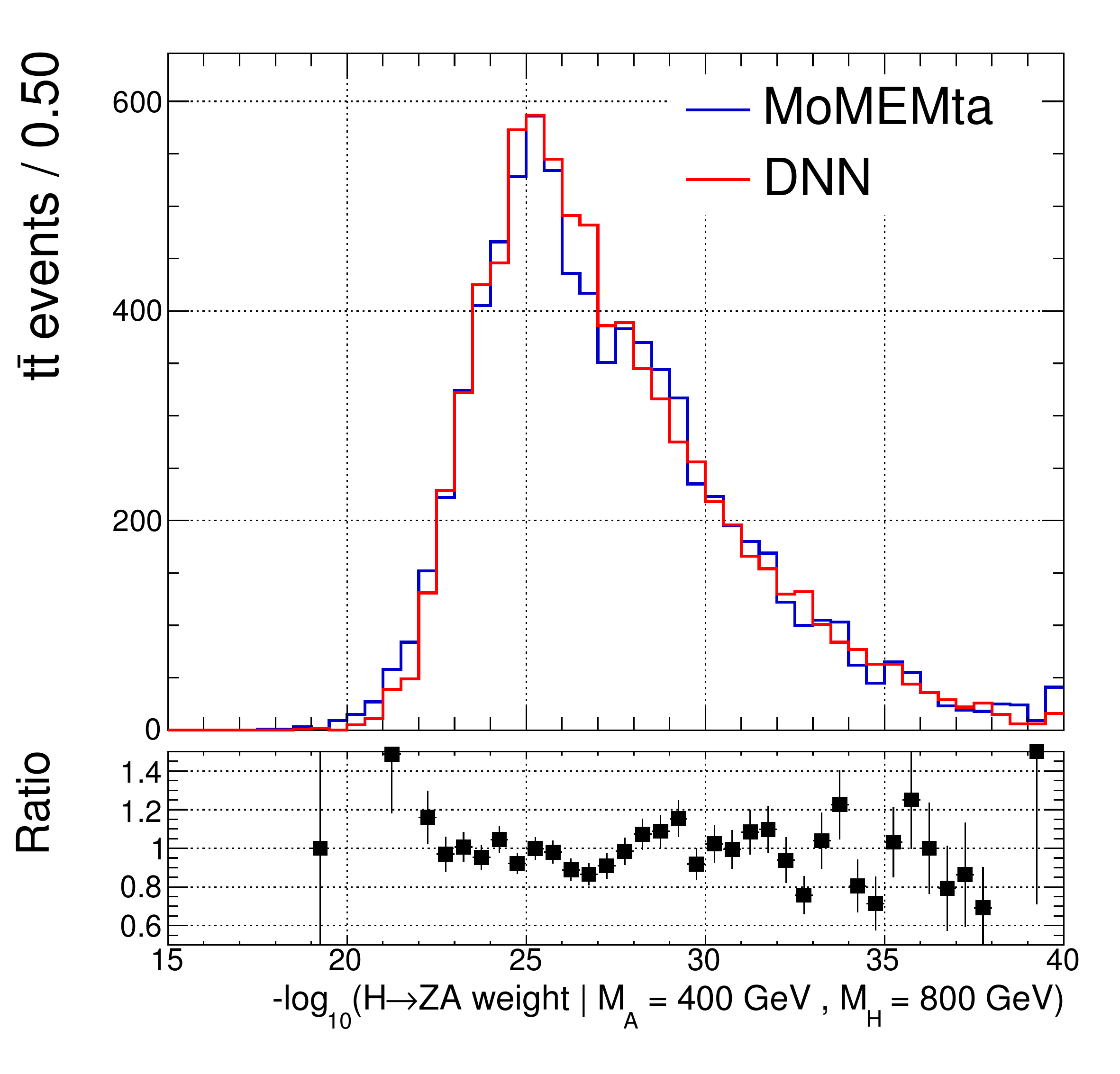}\hfill
	\includegraphics[width=.33\textwidth]{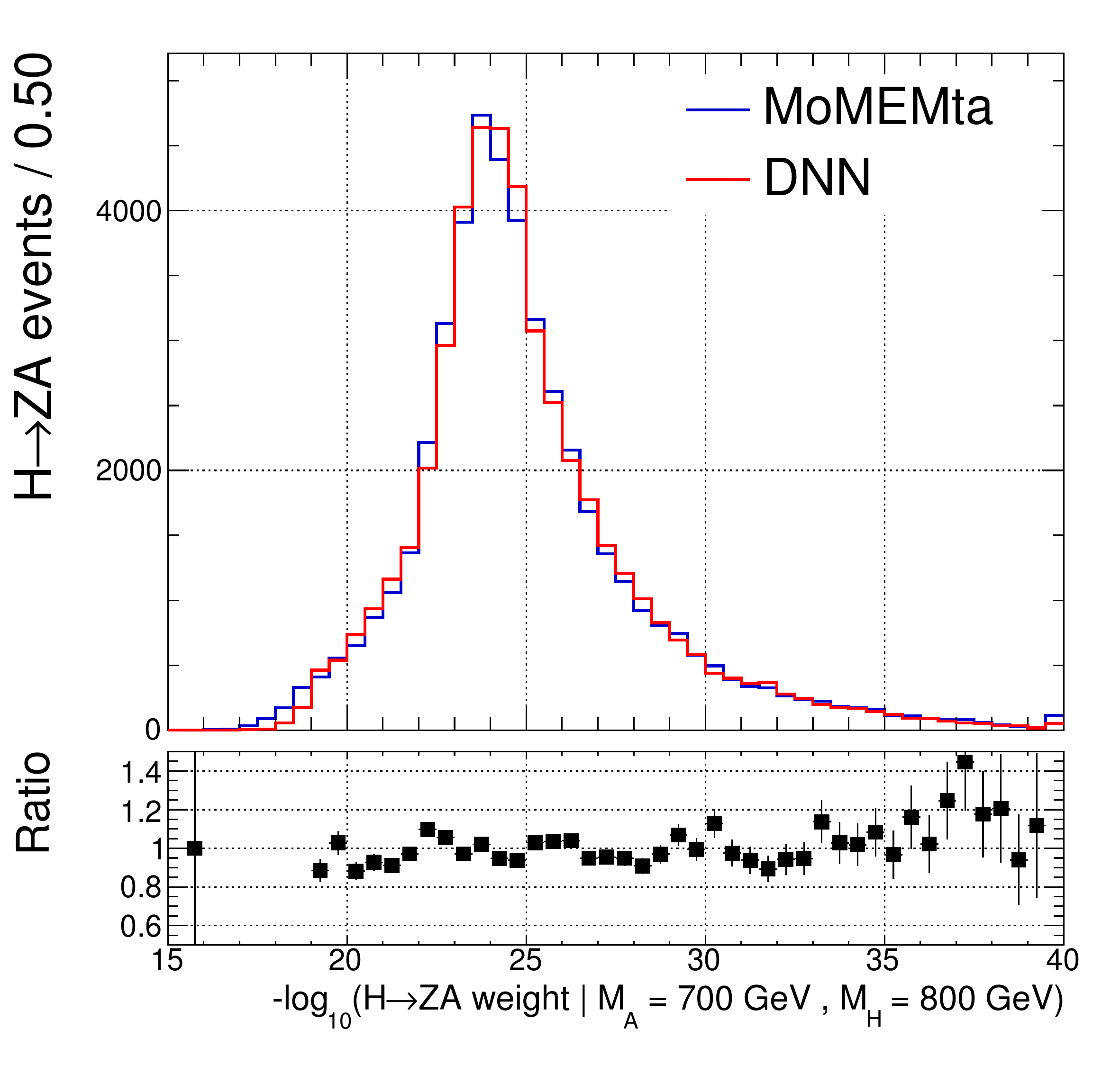}
	\caption{Distributions of the event information $I'_{H\to ZA}$ of the specific mass point ($M_A = \SI{400}{\GeV}$ , $M_H = \SI{800}{\GeV}$) from MoMEMta and the DNN for the three samples : Drell-Yan (left), $t\bar{t}$ (middle) and $H \rightarrow ZA$ (right) events.}
	\label{fig:signal_weight}
\end{figure}

From a pure technical point of view the best model for the $H \rightarrow ZA$ case is intrinsically similar to that of Drell-yan and $t\bar{t}$. It consists of 8 layers of 300 neurons with \textit{relu} and \textit{selu} activation functions for the hidden and output layers respectively, trained with a small L2 factor. It is however conceptually radically different. In addition to the particles inputs, it was also given the mass parameters for each $H \rightarrow ZA$ weight provided as target. It has been trained on each event 23 times, for different $M_H$, $M_A$ and target weight. Not only has the DNN learned about the relation between the kinematics and the weight, but also the dependence on the process parameters. The comparison between the weights from MoMEMta and the DNN is shown in Figure~\ref{fig:signal_weight} for a specific mass configuration ($M_H = \SI{800}{\GeV}$, $M_A = \SI{400}{\GeV}$). 

As expected, the $t\bar{t}$ events have small $H \rightarrow ZA$ weights when the mass parameters are high. This is the reciprocal of the fact that high mass signals have low $t\bar{t}$ weights. Drell-Yan events also depict the same behavior since they rarely produce particles with high momenta, which is typically the case for $H \rightarrow ZA$ with high mass resonances. 
Notice also that the higher weights for Drell-Yan events occur when the difference in mass hypothesis $M_H-M_A$ is large. In that case, the $Z$ boson acquires a large boost not commonly observed in Drell-Yan events.

\begin{figure}[htp]
	\centering
	\includegraphics[width=.33\textwidth]{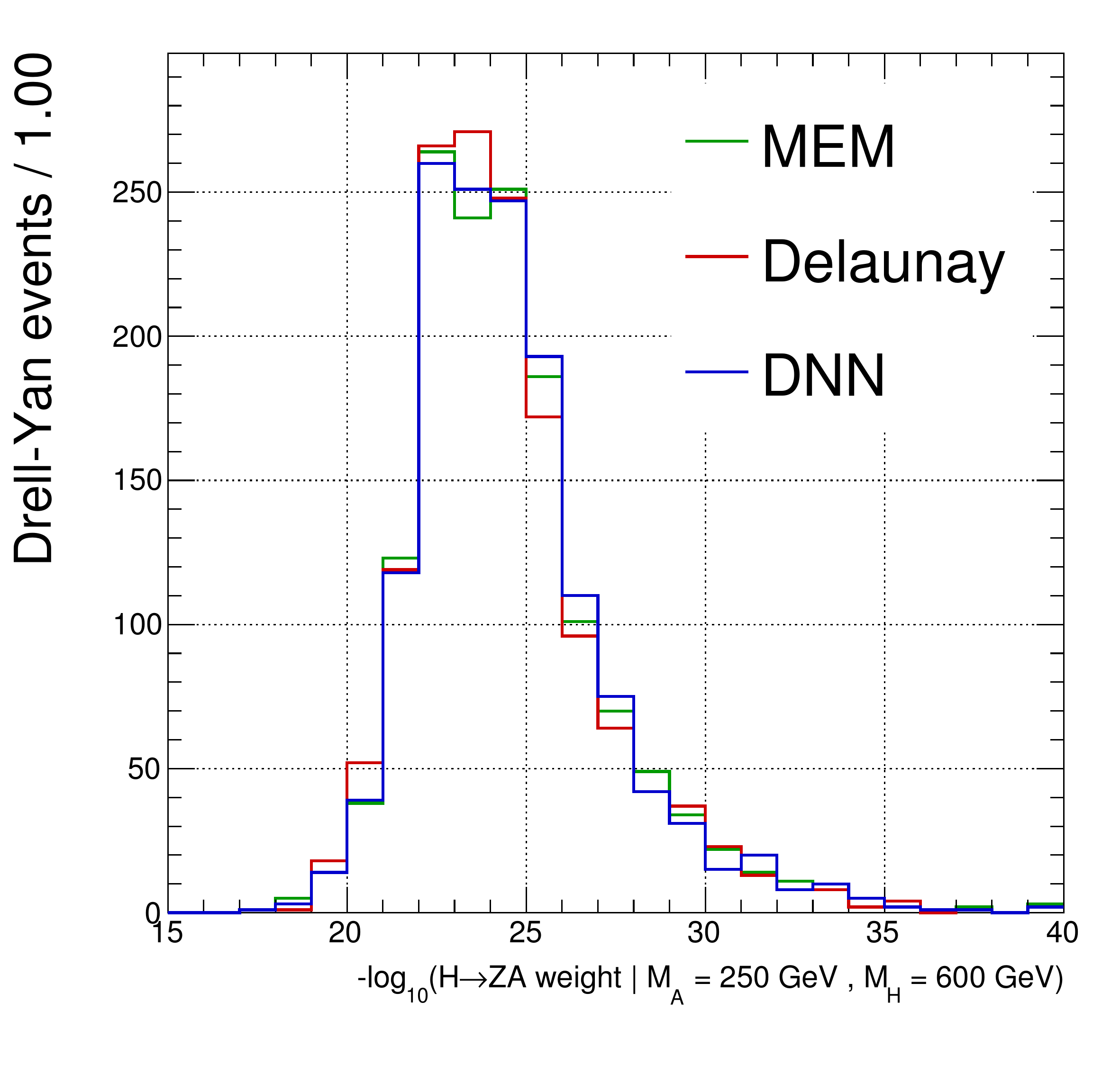}\hfill
	\includegraphics[width=.33\textwidth]{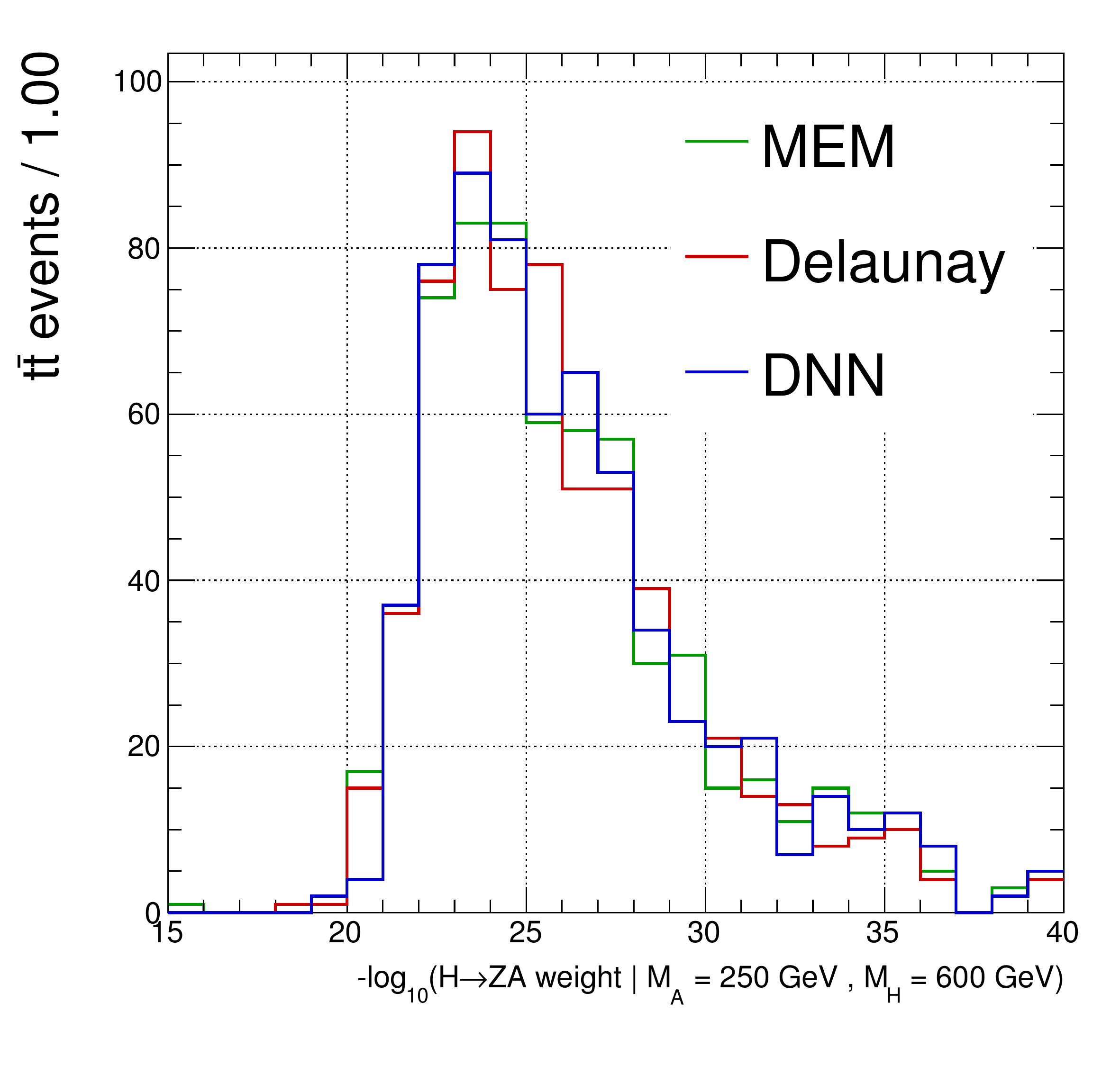}\hfill
	\includegraphics[width=.33\textwidth]{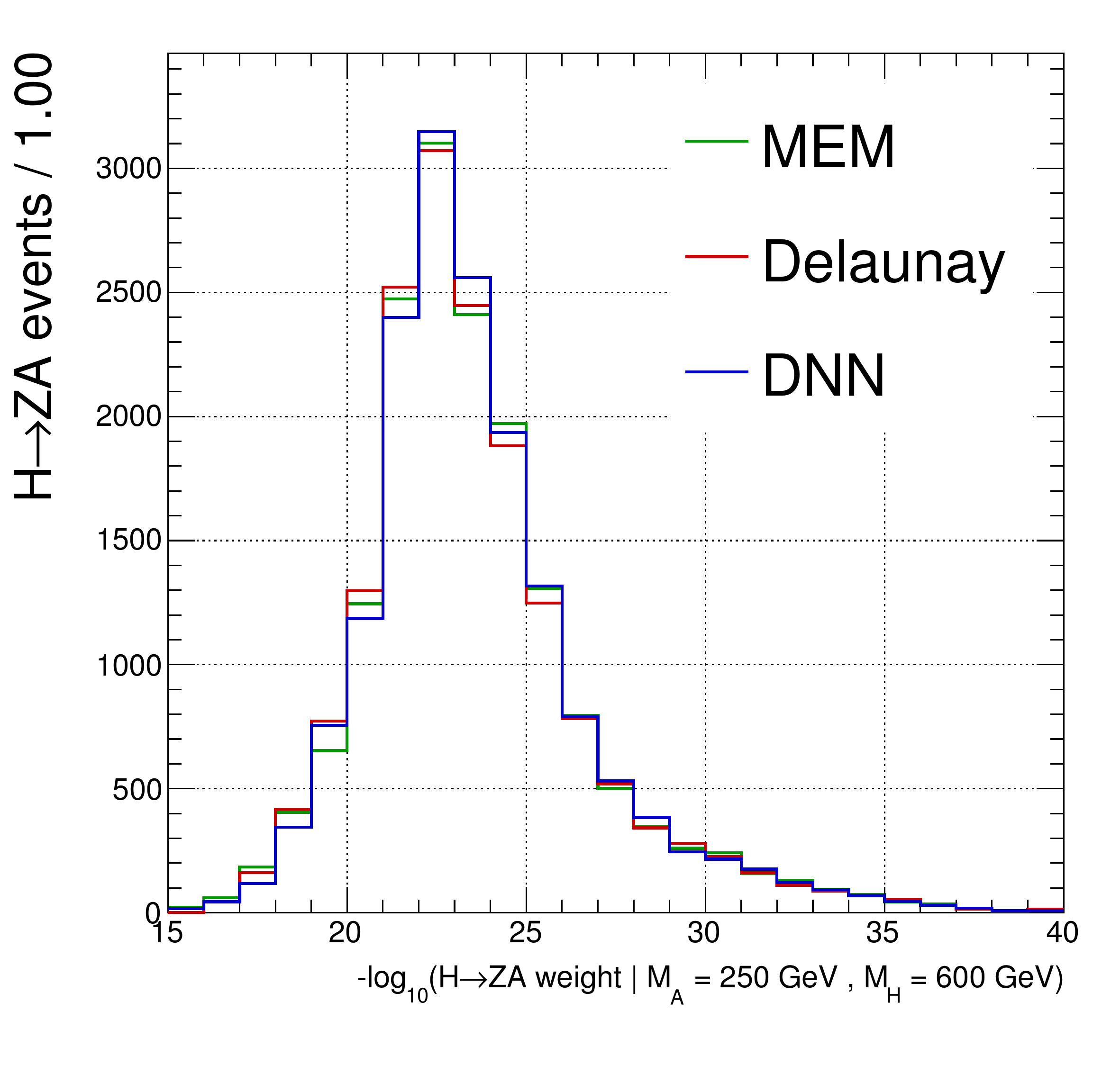}
	\caption{Distributions of the event information $I'_{H\to ZA}$ for the mass point ($M_A = \SI{250}{\GeV}$ , $M_H = \SI{600}{\GeV}$) not seen during the training. The true distribution from MoMEMta is in green, the Delaunay interpolation using the other weights is in red and the output of the DNN (not trained at this mass point) is in blue. The three samples used are Drell-Yan (left), $t\bar{t}$ (middle) and $H \rightarrow ZA$ (right) cases.}
	\label{fig:interpolation}
\end{figure}

In our specific case the parameter space is two-dimensional. To test the interpolation capabilities of the network, a new set of weights was computed with parameters $M_H = \SI{600}{\GeV}$ and $M_A = \SI{250}{\GeV}$ (never seen during the training) on a small subset of the initial samples (1K events per $H \rightarrow ZA$ sample, 5K events for the Drell-Yan and $t\bar{t}$ samples). For reference, the Delaunay technique --- a piecewise-linear interpolation --- was employed to obtain the weights at these parameters values from closeby computed points. 
This is compared to the DNN applied to these events without retraining in Figure~\ref{fig:interpolation}.
Both method perform equally well, demonstrating that the DNN is properly interpolating the parameter space from observed samples. 
Note that while the Delaunay technique is relatively fast, the main bottleneck is that it requires some points to start from, which means that each event still needs to be computed for several mass points with MoMEMta. In addition the granularity will still scale exponentially with the parameter space dimension. On the contrary, there is no need to use MoMEMta anymore once the DNN is trained and while the two methods give the same result, the DNN can be orders of magnitude faster -  especially in multi-dimensional parameter space.

\begin{table}[htp]
	\caption{Summary of evaluation and training times, and DNN topology. The topology includes the number of layers and neurons per layers as well as the L2 regularization value, the activation functions are always relu and selu for the hidden and output layers. Additionally all DNN have been trained with a batch size of 512 and initial learning rate of 0.001. A learning rate scheduler and early stopping were implemented to stop the training at the validation loss curve plateau before the 100 epochs limit (even much sooner than that for the $H \rightarrow ZA$ case). The training size for the $H \rightarrow ZA$ DNN takes into account that each event is seen for each mass configuration.}
	\centering
	\resizebox{\textwidth}{!}{  
	\begin{tabular}{|l|c|c|c|c|c|c|c|c|}
		\hline
		\multirow{2}{*}{MEM Hypothesis} & \multirow{2}{*}{MoMEMta evaluation time} & \multicolumn{3}{c}{DNN topology} \vline & \multicolumn{2}{c}{DNN training time} \vline & \multirow{2}{*}{DNN evaluation time} \\
		\cline{3-7}
								 &  					                    & $N_{layers}$ & $N_{neurons}$ & L2 & Training size & Time / epoch & \\
		\hline
		Drell-Yan & \SI{3.6}{\second} / event & 6 & 200  & 0 & 800K events & \SI{5}{\minute} &  \SI{110}{\micro\second}  / weight \\
		\hline
		$t\bar{t}$ &  \SI{12}{\second} / event & 8 & 500 & 0.1 & 800K events &  \SI{10}{\minute} & \SI{150}{\micro\second}  / weight \\
		\hline
		$H \rightarrow ZA$ &  \SI{600}{\second} / event & 8 & 300 & 0.1 & 5,7M events &  \SI{40}{\minute} & \SI{120}{\micro\second}  / weight / parameter \\
		\hline
	\end{tabular}}
	\label{tab:DNN_comp}
\end{table}

\subsection{Applications and studies}
\label{applications}
While the weight distributions presented so far are a good indicator to evaluate the regression, they do not bring information about the event-by-event agreement. It is also difficult to evaluate if the residual difference is physically relevant.
In the following, we will look at typical applications of the MEM in analysis, with the Information used as an input of the MVA discriminant, or interpreted as a likelihood.
This will allow also to better understand the status of the invalid weights and the sensitivity to systematic uncertainties.

\subsubsection{Discriminant} 
\label{sec:discriminant} 
A very simple discriminant between two hypotheses $\alpha$ and $\beta$ can be defined as 
\begin{equation}
	\mathcal{D}(x) = \frac{P(x|\alpha)}{P(x|\alpha) + P(x|\beta)} = \frac{W(x|\alpha)}{W(x|\alpha) + \gamma W(x|\beta)} \textrm{  where  } \gamma = \frac{\sigma^{vis}_{\beta}}{\sigma^{vis}_{\alpha}}.
	\label{eqn:discriminant}
\end{equation} 
The discriminant can be close to one or zero depending on which hypothesis $\alpha$ or $\beta$ prevails (respectively). 
For illustration, $\alpha$ and $\beta$ can be taken to be respectively the $t\bar{t}$ and Drell-Yan processes. As an evaluation criterion for this discriminant we have used the Receiver Operating Characteristic (ROC) curve. Although it does not impact the ROC curve, the shape of the discriminant will be impacted by the factor $\gamma$ in the denominator, we have arbitrarily taken here $\gamma = 1$.

\begin{figure}[htp]
	\centering
	\includegraphics[width=.6\textwidth]{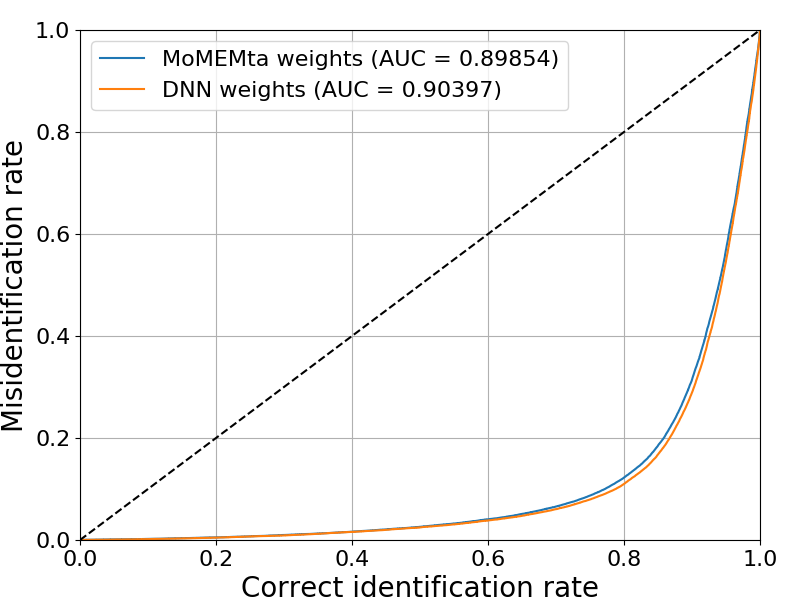}
	\caption{ROC curve of the discriminant. The x axis represents the probability for an event to be classified in the correct process ($t\bar{t}$ or Drell-Yan) while the y-axis represents the misidentification probability. The AUC score is the area under curve.}
	\label{fig:ROC}
\end{figure}

The ROC curves obtained with the weights coming from both the integration in the MEM and from the DNN are shown in Figure~\ref{fig:ROC}. The weights produced by the DNN actually provide a slightly better discriminant than the ones from MoMEMta. 
The difference can be traced to outliers present in the MoMEMta calculation, while the DNN behavior is smoother by nature and has fewer of them.

The discriminant in Equation~\ref{eqn:discriminant} is limited to a classification between two hypotheses which is a bit restrictive. In addition its simple definition might not make use of the full information encapsulated in the MEM weights. A discriminant for higher dimension parameter space could be generalized but is not guaranteed to be optimal. In this paper we decided to follow a different path by using a classifier based on the MEM weights and leave to it the determination of an optimal decision function. A natural choice here is the use of a classifying DNN with three output nodes whose inputs are the weights for the three processes under study. The DNN is trained to maximize the probability of correct identification using a binary cross-entropy loss function. There are two ways one can define the classifiers inputs based on the parametric definition of the $H \rightarrow ZA$ weights. A \textit{global} classifier is used to find an excess in the whole mass plane, regardless of its location. In the spirit of the analysis in Reference~\cite{HZA} and anticipating the search for a specific resonance, a \textit{parametric} classifier is given the knowledge of the mass plane point of interest and can therefore be used to find an excess at a given place. On the one hand the global classifier is less sensitive because the excess needs to be large to be noticeable on the whole plane, on the other hand there is no need to correct for the look-elsewhere effect which would be the case for the parametric classifier that is evaluated at several locations.

\begin{figure}[htp]
	\centering
	\includegraphics[width=.33\textwidth]{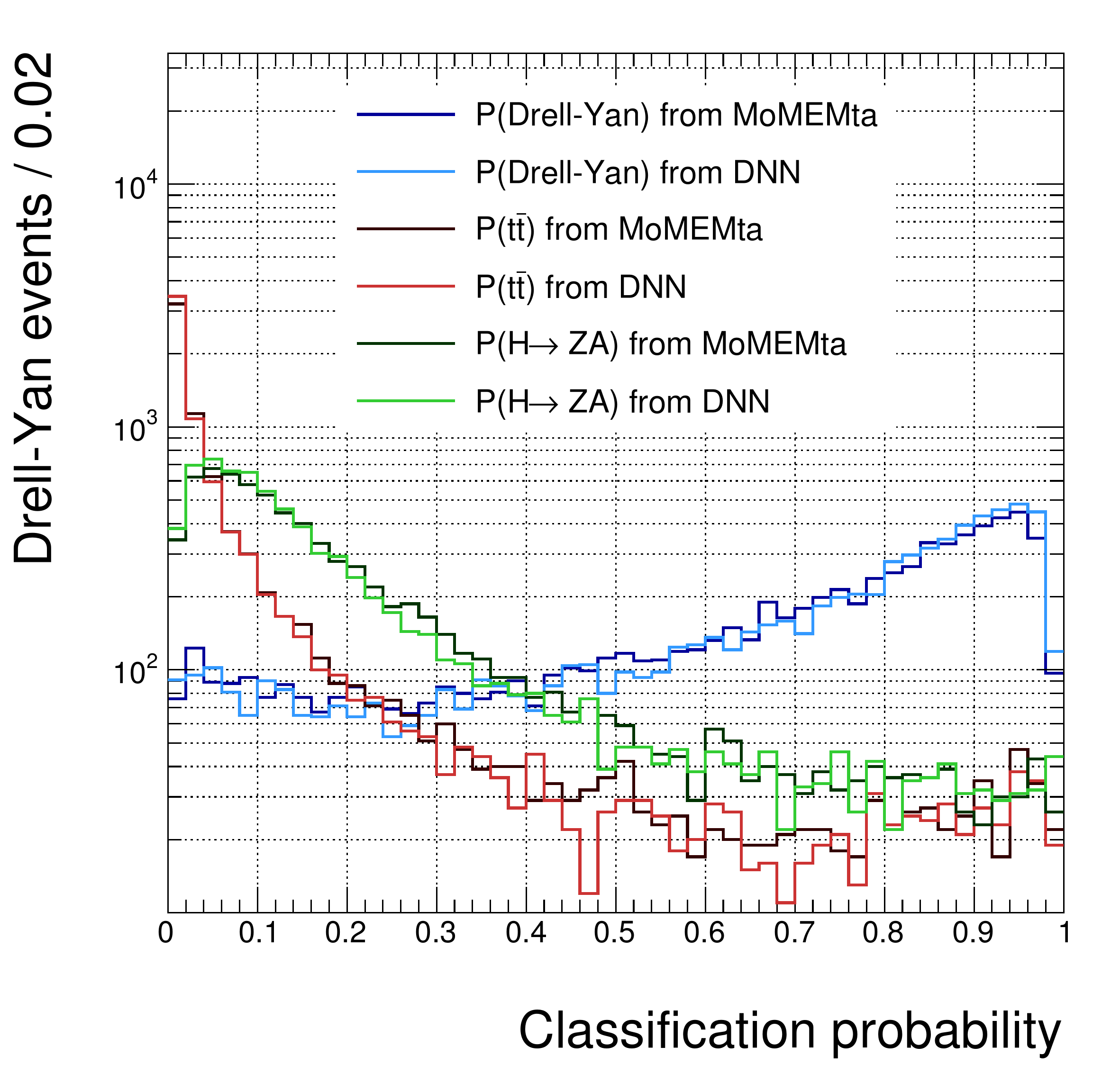}\hfill
	\includegraphics[width=.33\textwidth]{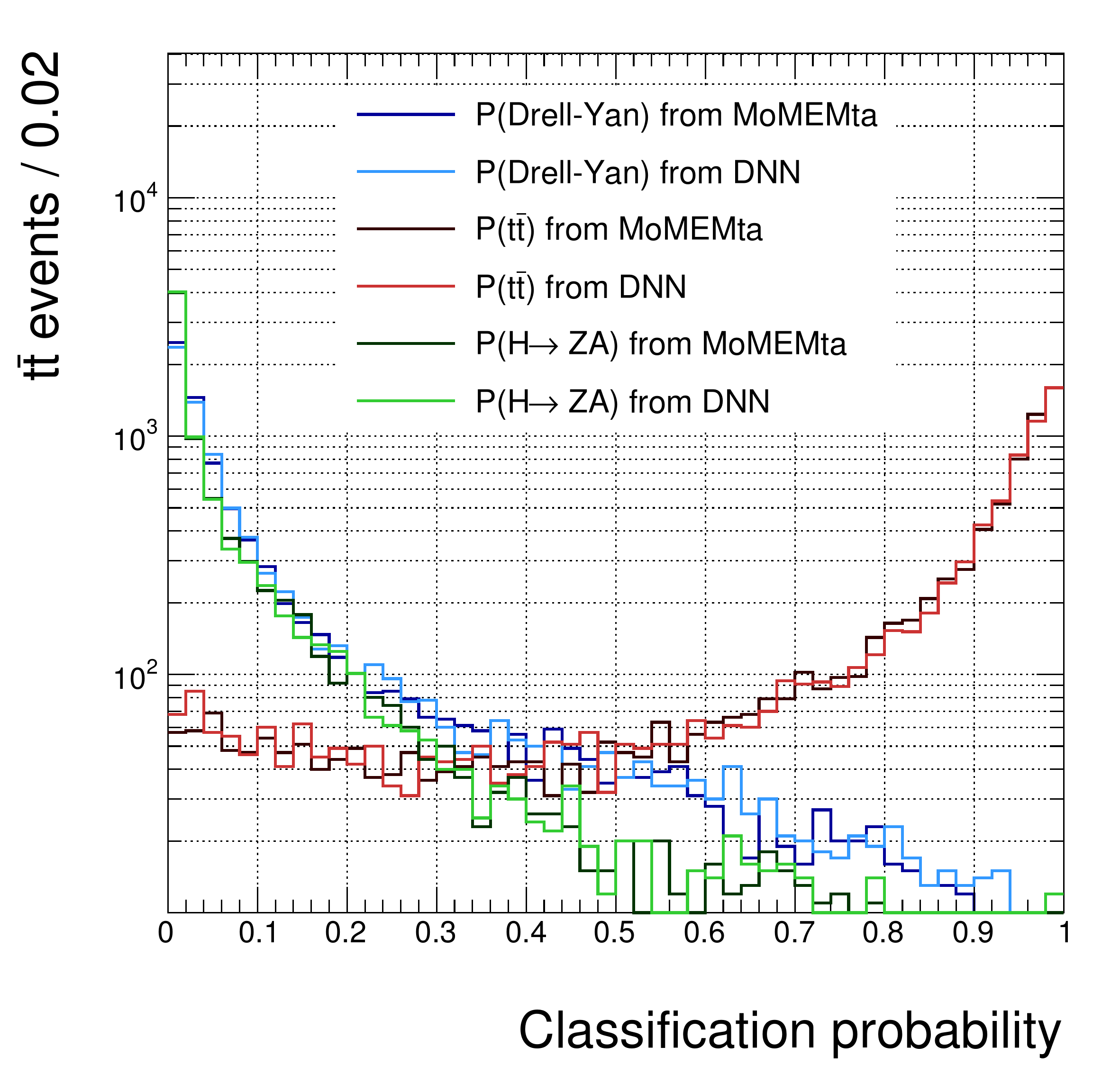}\hfill
	\includegraphics[width=.33\textwidth]{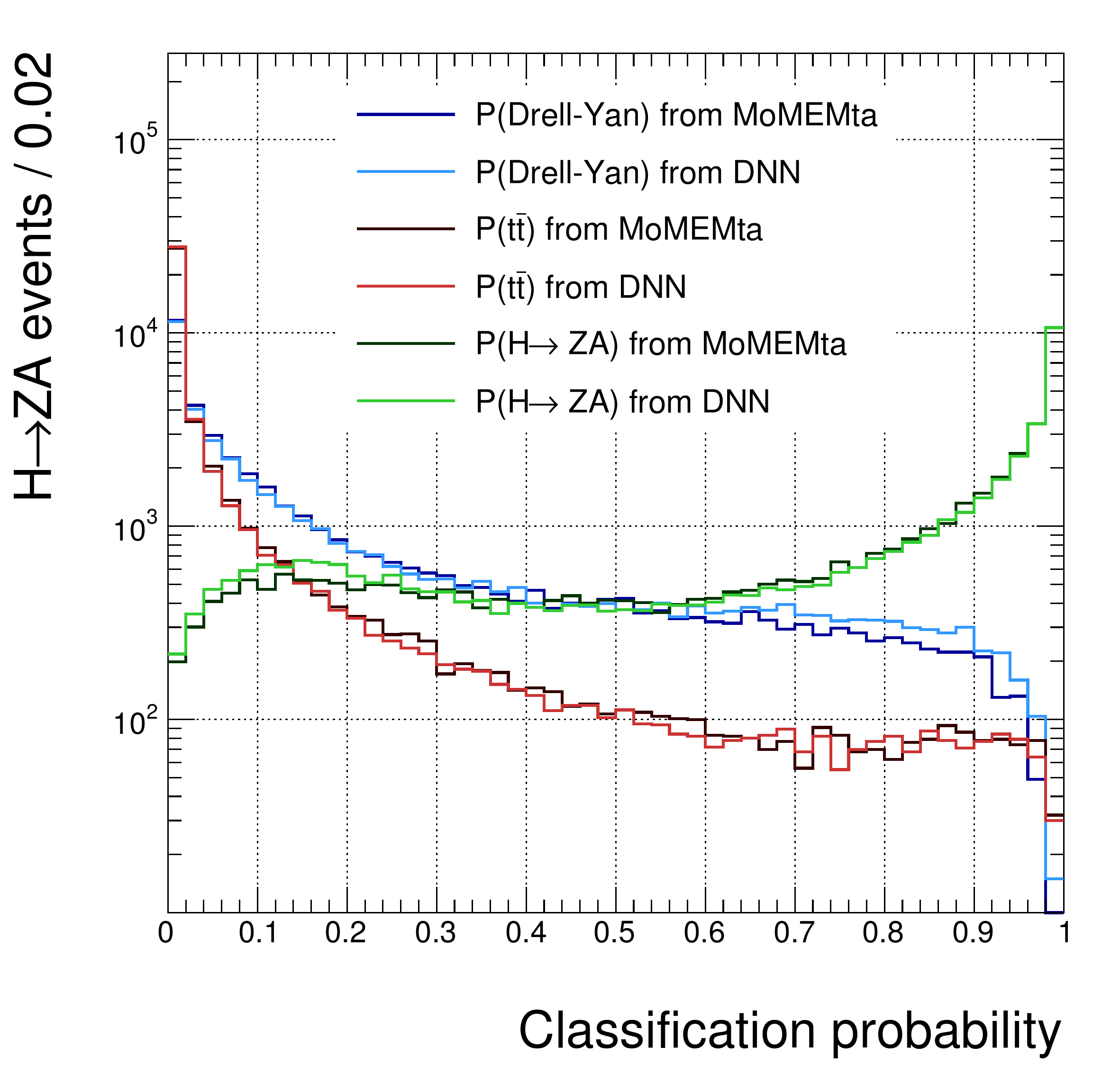}
	\caption{Distributions of the class probabilities with the global classifier applied on weights coming from both MoMEMta and the DNN for the three samples : Drell-Yan (left), $t\bar{t}$ (middle) and $H \rightarrow ZA$ (right) events.}
	\label{fig:prob}
\end{figure}

\begin{figure}[htp]
	\centering
	\begin{subfigure}[b]{0.32\linewidth}
		\centering\includegraphics[width=\linewidth]{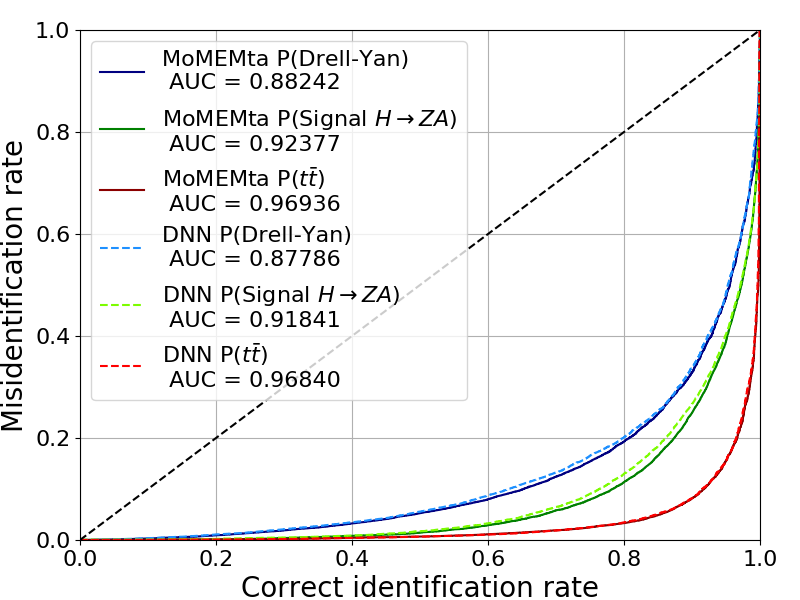}
		\caption{\label{fig:MultiROC_global}}
	\end{subfigure}
	\begin{subfigure}[b]{0.32\linewidth}
		\centering\includegraphics[width=\linewidth]{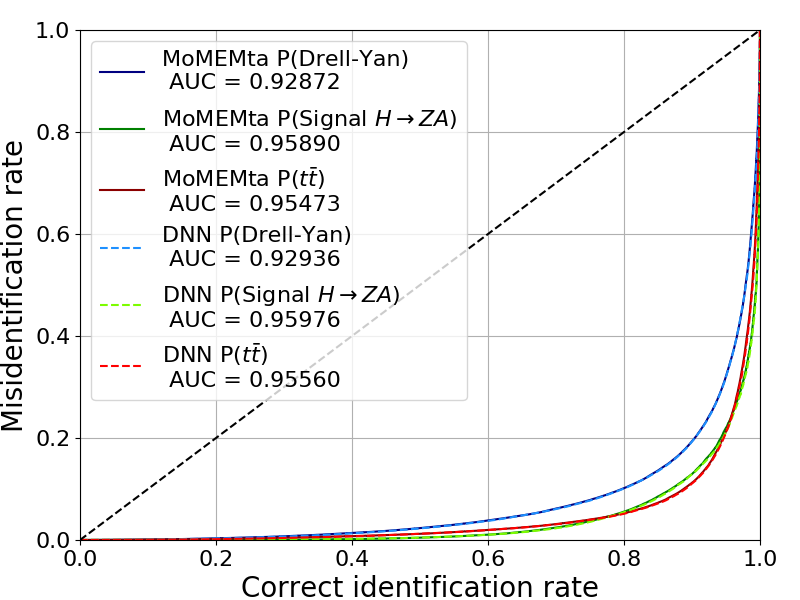}
		\caption{\label{fig:MultiROC_param}}
	\end{subfigure}
	\begin{subfigure}[b]{0.32\linewidth}
		\centering\includegraphics[width=\linewidth]{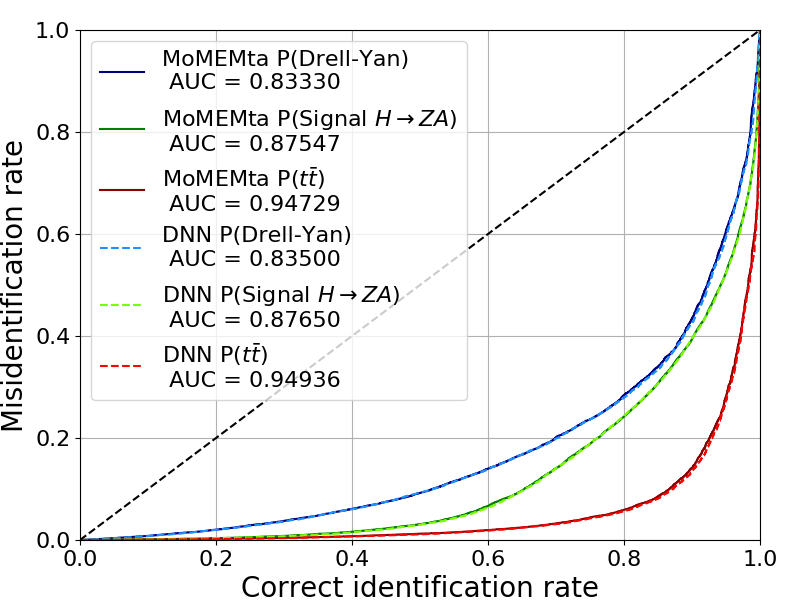}
		\caption{\label{fig:MultiROC_back}}
	\end{subfigure}%
	\caption{ROC curves of the Global (a), Parametric (b) and Simple (c) classifiers. 
	The fraction of events incorrectly classified as coming from a given process is represented as a function of the probability to be correctly identified as such for each of the three processes.
	The ROC curves are given for both the weights from MoMEMta (solid lines) and the ones coming from the DNN (dashed line). The AUC score is the area under each curve.}
	\label{fig:MultiROC}
\end{figure}

The inputs of the global classifier are the Drell-Yan, $t\bar{t}$ and the 23 $H \rightarrow ZA$ weights. As we have no knowledge of the actual value ot the masses during the training on simulated events we need a good enough coverage of the parameter space. At least one input should provide sensitivity to a given hypothesis.
The classification probabilities are given in Figure~\ref{fig:prob} and the corresponding ROC curves are compared in Figure~\ref{fig:MultiROC_global} using both the weights from MoMEMta and the regressive DNNs.

\begin{figure}[htp]
	\centering
	\includegraphics[width=\textwidth]{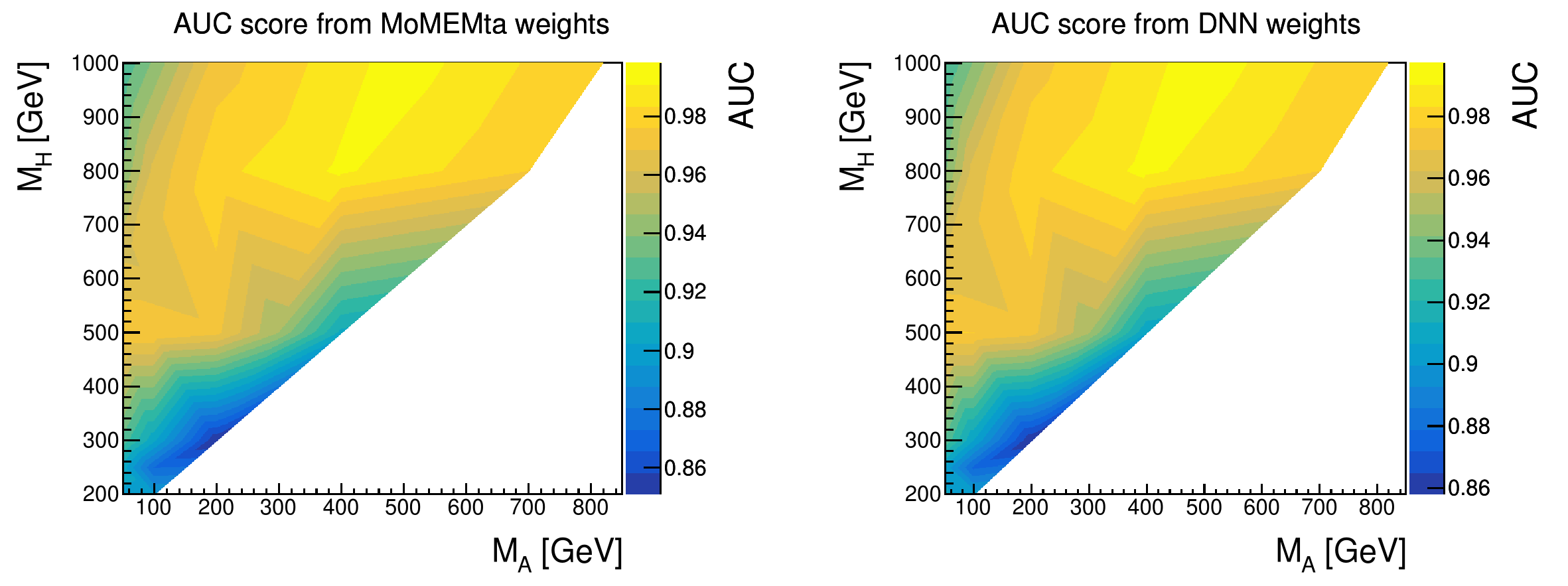}
	\caption{Distribution of the AUC score of the $H \rightarrow ZA$ classification for the weights coming from MoMEMta (left) and the DNN (right).}
	\label{fig:AUC_map}
\end{figure}

The parametric classifier also takes as inputs the Drell-Yan and $t\bar{t}$ weights but only one $H \rightarrow ZA$ weight with the corresponding $M_A$ and $M_H$ parameters. 
For $H \rightarrow ZA$ events, the actual parameter value is used, while for Drell-Yan and $t\bar{t}$ events they can either be attributed a random parameter point --- in the same proportions as in $H \rightarrow ZA$ events --- or repeated for every parameter point found in the $H \rightarrow ZA$ events.
The latter was used to artificially increase the statistics. The associated ROC curves are shown in Figure~\ref{fig:MultiROC_param}, averaging over all the mass points. 
The dependence on the performance as a function of these mass points is illustrated by the AUC score in Figure~\ref{fig:AUC_map}. 
Naturally, the best performance is achieved away from the regions heavily populated by other processes.

As a comparison, a simple classifier ROC curve with only the Drell-Yan and $t\bar{t}$ weights as inputs is shown in Figure~\ref{fig:MultiROC_back}. Even though this simple classifier is suboptimal for $H \rightarrow ZA$, it reaches reasonable performance. Additionally the Drell-Yan and $t\bar{t}$ classifications improve when provided with the $H \rightarrow ZA$ weight information. The MEM weights can provide discriminating power even for processes other that the one used in its computation.  

All the classifiers are trained with weights from MoMEMta and the ROC curves shown in Figure~\ref{fig:MultiROC} are evaluated with weights from both methods. The regression errors introduced when using the regressive DNNs are propagated through the classifiers but the loss in performance is negligible. Only in the global classifier can the MEM and DNN curves be distinguished due to the residual differences already highlighted in Figure~\ref{fig:signal_weight} that add up for all the $H \rightarrow ZA$ weights inputs.

\subsubsection{Invalid weights}

\begin{figure}[htp]
	\centering
	\includegraphics[width=.33\textwidth]{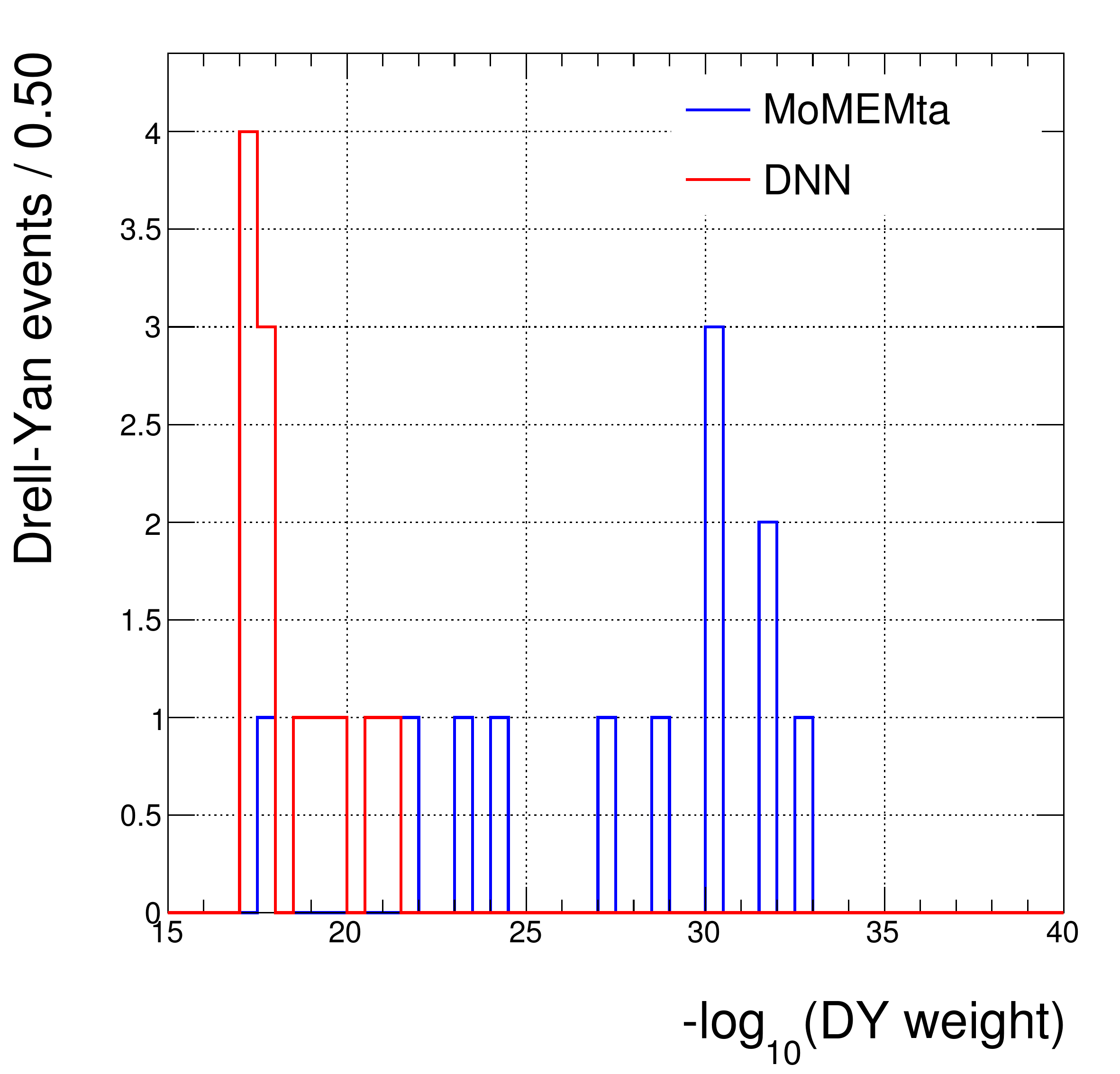}\hfill
	\includegraphics[width=.33\textwidth]{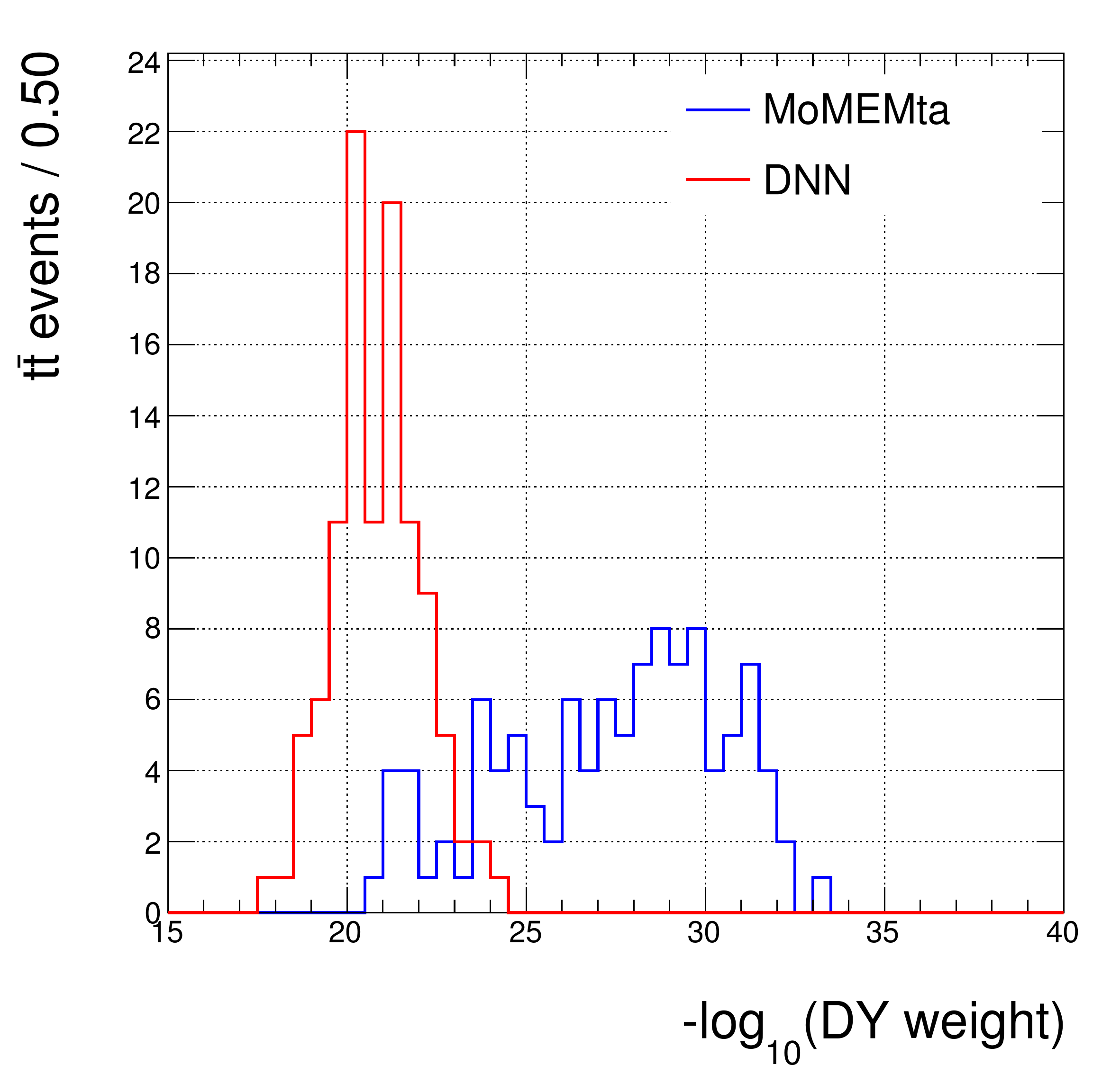}\hfill
	\includegraphics[width=.33\textwidth]{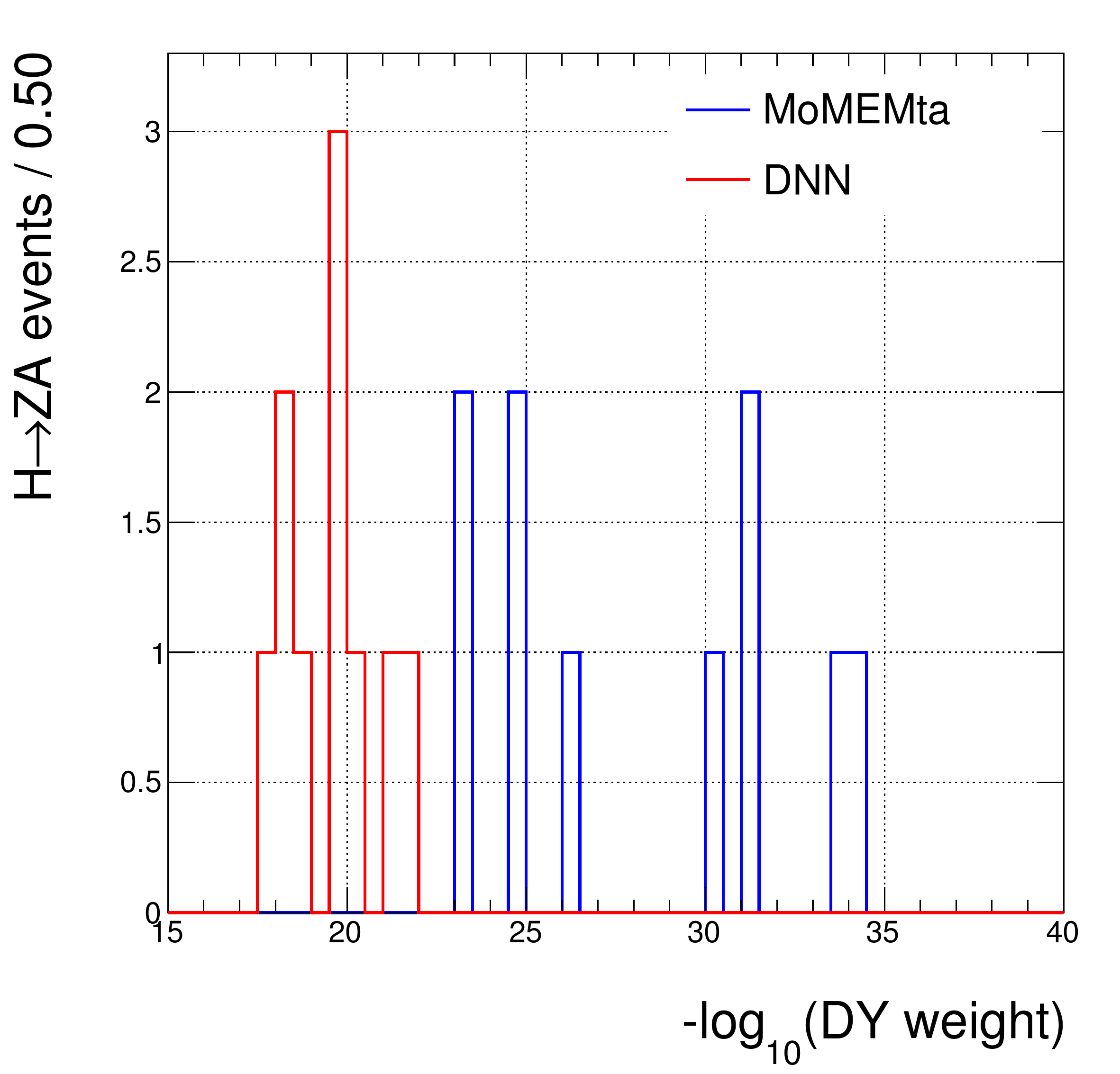}
	\caption{Distributions of the event information $I'_{DY}$ for events where the integration initially failed either from MoMEMta (blue) or the DNN trained only on valid weights (red) for the three samples : Drell-Yan (left), $t\bar{t}$ (middle) and $H \rightarrow ZA$ (right) events.}
	\label{fig:invalid_DY}
\end{figure}

\begin{figure}[htp]
	\centering
	\includegraphics[width=.33\textwidth]{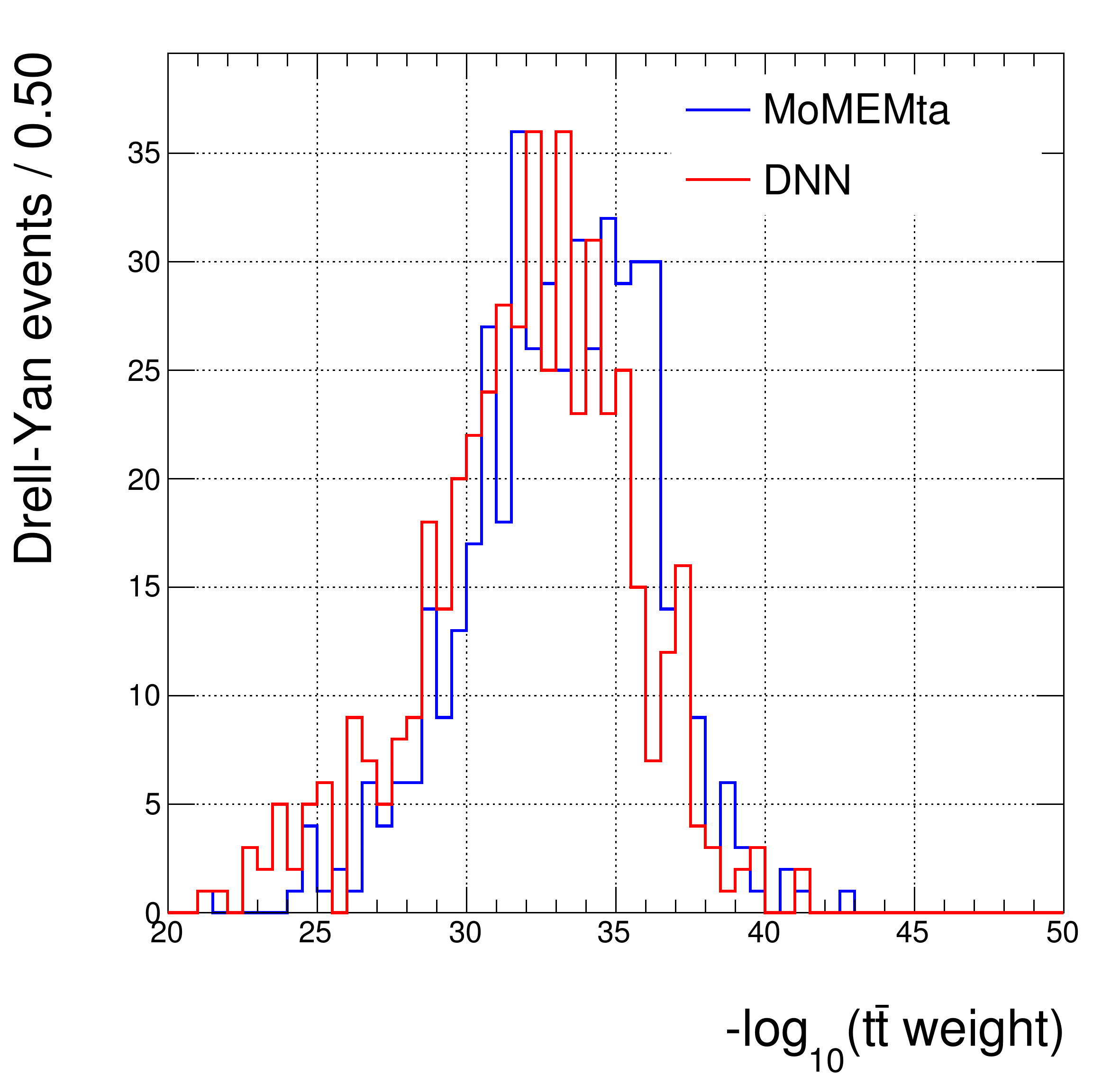}\hfill
	\includegraphics[width=.33\textwidth]{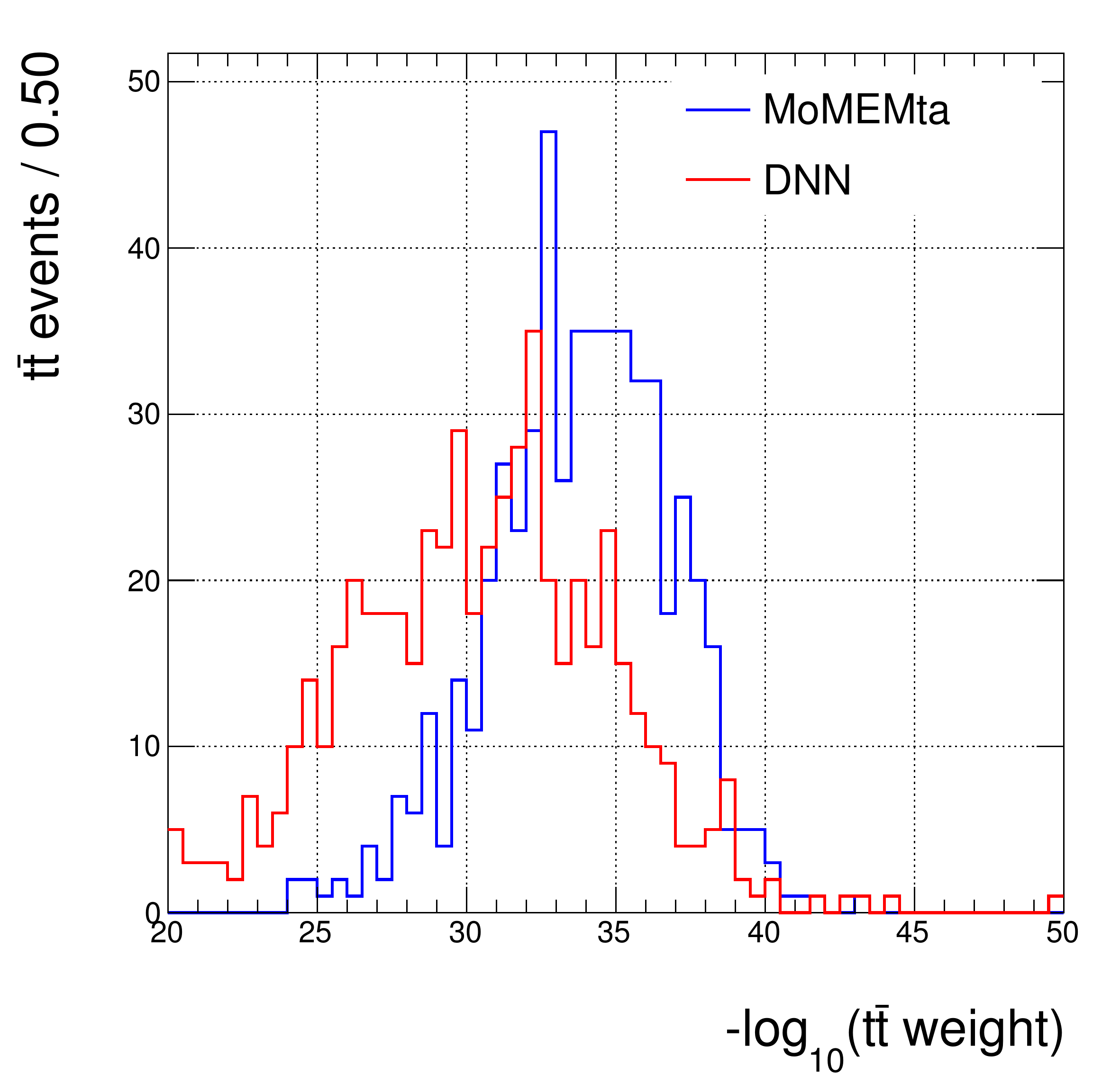}\hfill
	\includegraphics[width=.33\textwidth]{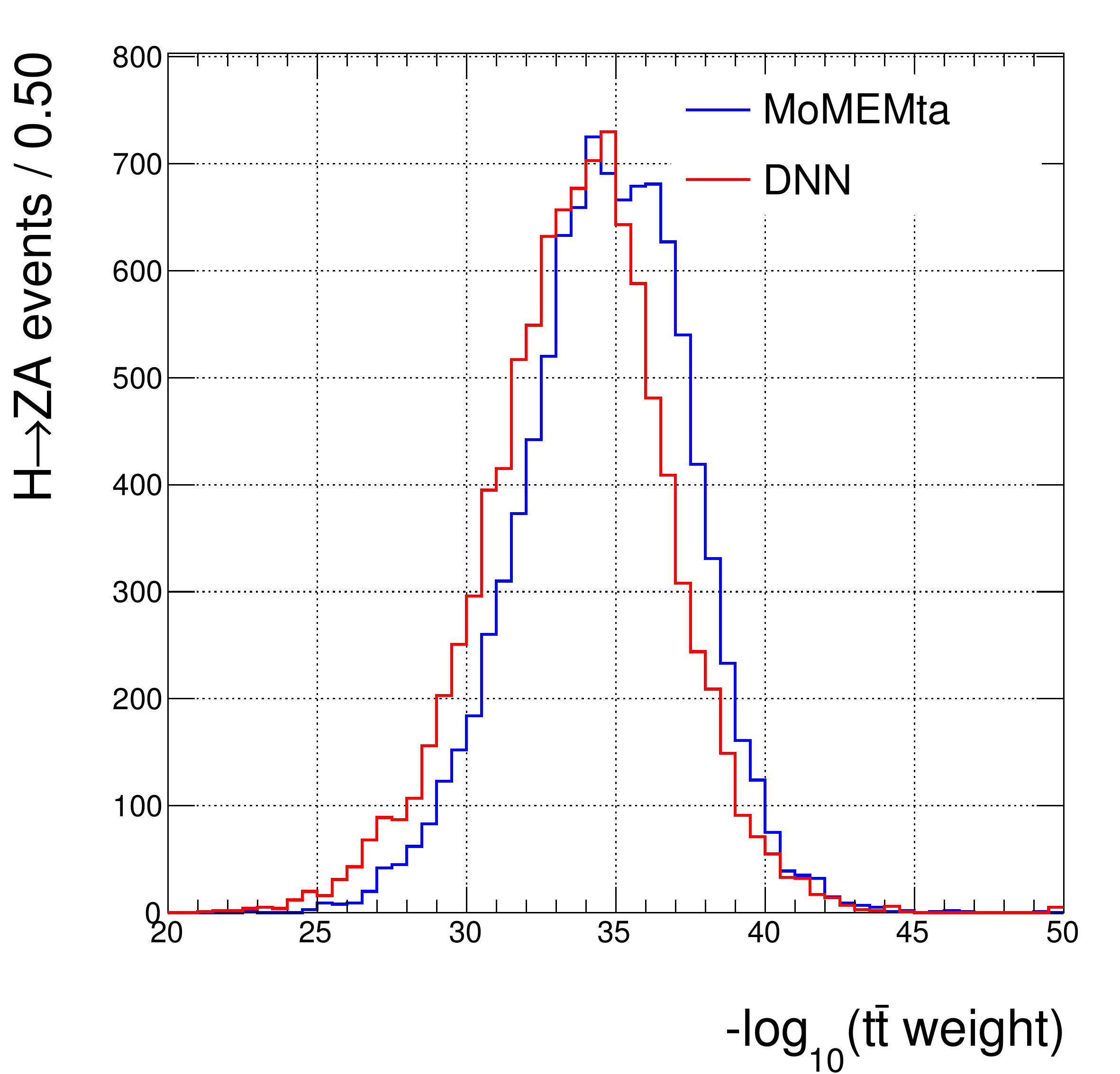}
	\caption{Distributions of the event information $I'_{t\bar{t}}$ for events where the integration initially failed either from MoMEMta (blue) or the DNN trained only on valid weights (red) for the three samples : Drell-Yan (left), $t\bar{t}$ (middle) and $H \rightarrow ZA$ (right) events.}
	\label{fig:invalid_TT}
\end{figure}

As discussed earlier, the MoMEMta integration may fail and result in "invalid weights". It is nevertheless trivial to evaluate the DNN for the corresponding events. The result can be compared to the output of MoMEMta when the integration is made to converge by increasing the number of sampling points, as done in Figures~\ref{fig:invalid_DY} and~\ref{fig:invalid_TT}, respectively for Drell-Yan and $t\bar{t}$ weights. 
For both cases it is obvious that the DNN does not agree with MoMEMta for these events with invalid weights.
In the case of Drell-Yan, the DNN seems to deliver values similar to what is obtained for normal events, while MoMEMta returns small probabilities even after allowing more iterations. 
The picture is quite different for the $t\bar{t}$ case, where the network returns consistently smaller weights (even though these events were never seen during training).
The question of what is happening with these events and whether we can trust MoMEMta with these very small weights remains open at this stage.

\begin{figure}[htp]
	\centering
	\includegraphics[width=.49\textwidth]{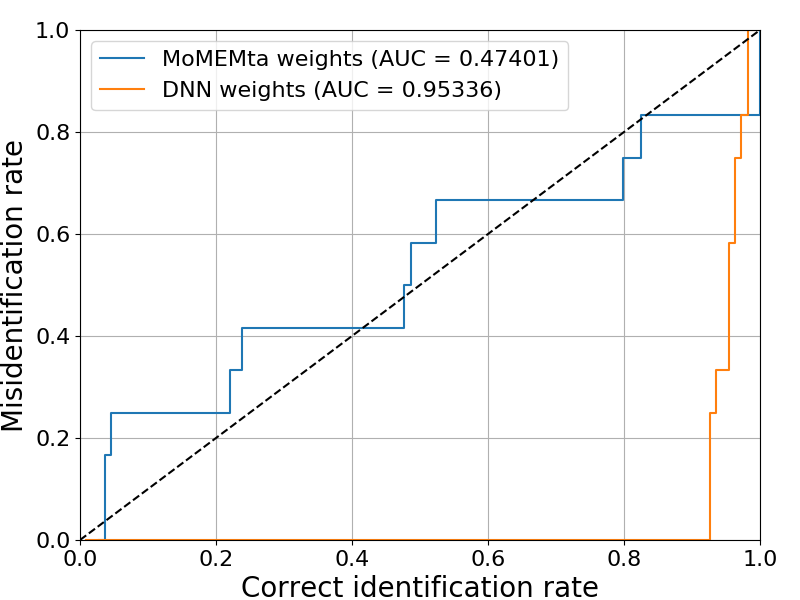} \hfill
	\includegraphics[width=.49\textwidth]{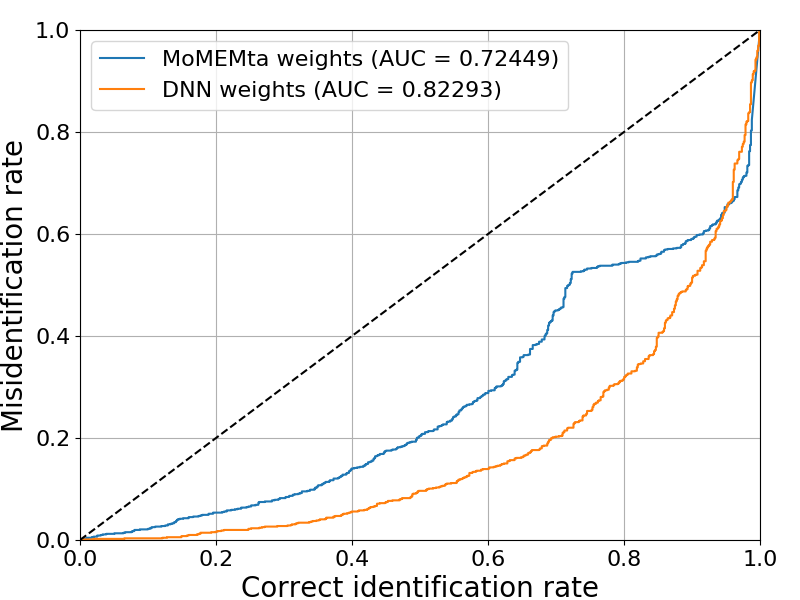}
	\caption{ROC curve of the discriminant for two cases : the invalid Drell-Yan weights (left) and the invalid $t\bar{t}$ weights (right). In each case two ROC curves are displayed, one for each discriminant based on the methods with MoMEMta and the DNN.}
	\label{fig:cases_ROC}
\end{figure}

Applying the discriminant from Equation~\ref{eqn:discriminant} to compare the invalid weights computed with MoMEMta and the DNN, much better performance is obtained with the DNN inputs than with the MoMEMta inputs (Figure~\ref{fig:cases_ROC}). This may suggest that the DNN provides a more reliable information than MoMEMta for these events, but more studies are needed to get a conclusive answer.


\subsubsection{Effect of the systematics}

Already heavy in terms of computing in its simplest form, the MEM becomes even more demanding when the effect of systematic uncertainties has to be evaluated.
Indeed, the effect of uncertainties affecting the event kinematics cannot be propagated without recomputing the matrix element integral, with relatively few opportunities to optimize this calculation.
While the DNN ansatz proposed here alleviates the impact of this additional integration, it is important to verify that the regression performed during the training phase is robust against these systematic effects too.

As an example, we will look at the jet energy scale (JES), which is potentially among the most dangerous effects, due to the rather poor resolution of jets compared to leptons in hadron collider experiments.
We will not consider the impact of the jet energy resolution, which is mostly covered by the transfer functions.
To emulate a potential JES we have applied an upward scaling of the jet energy by 10\% --- which is an extreme case --- for each of the bjets on a subset of the events and computed the corresponding new weights both with MoMEMta and with the existing DNN parameterization (thus without retraining).

\begin{figure}[htp]
	\centering
	\begin{subfigure}[b]{0.42\linewidth}
		\includegraphics[width=\textwidth]{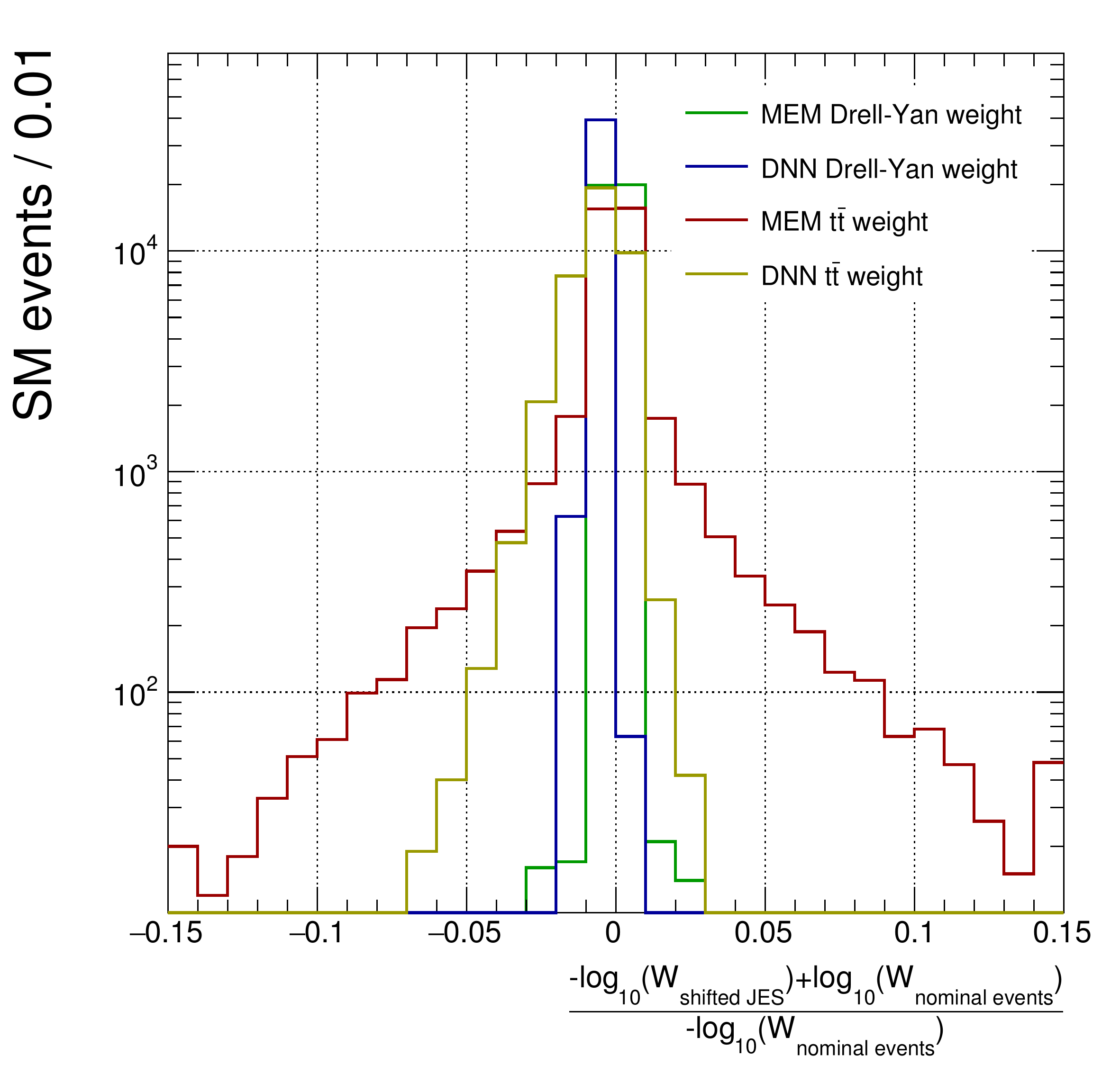}
		\caption{\label{fig:JES_weights}}
	\end{subfigure}
	 \begin{subfigure}[b]{0.53\linewidth}
	 	\includegraphics[width=\textwidth]{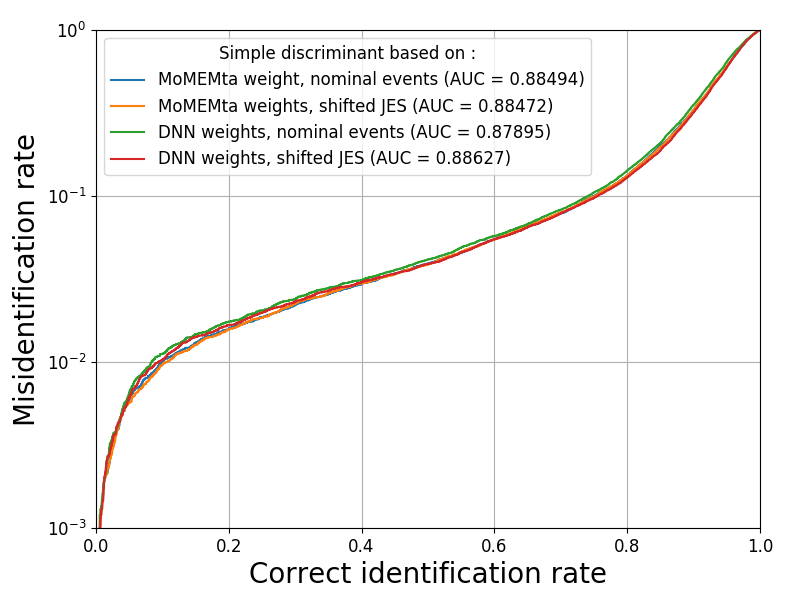}
			\caption{\label{fig:JES_ROC}}
	 \end{subfigure}
	\caption{Comparison of the effect of the JES corrections on the background weights produced by MoMEMta and the DNN. The left figure illustrates the change in event information when the JES corrections are applied for both weights and both methods. The right figure is the ROC curve of the discriminant based on these weights.}
	\label{fig:JES_comparison}
\end{figure}

The comparison of the weights when the JES shift is applied or not is shown in Figure~\ref{fig:JES_weights}. The regression error itself is not significantly impacted, as can be seen from the DNN bias and resolution in Table~\ref{tab:JESimpact}. This shows that the DNN is able to properly handle these modified events.

In addition, the discriminant from Section~\ref{sec:discriminant} has been evaluated using both MoMEMta and DNN inputs for nominal and modified events. 
The associated ROC curves are shown in Figure~\ref{fig:JES_ROC}.
This specific discriminant appears to be robust against variations of the jet energy scale, both in its traditional implementation and when using the DNN ansatz as input.

\begin{table}[htp]
	\caption{Regression bias and resolution in the event information when replacing the integration with MoMEMta by the DNN ansatz for the two SM weights with nominal and shifted JES events.}
	\centering
	\begin{tabular}{|c|c|c|}
		\hline
		& Regression bias & Regression resolution  \\
		\hline
		Nominal Drell-Yan & -0.1243 & 0.1383 \\
		\hline
		Shifted JES Drell-Yan & 0.0049 & 0.1351  \\
		\hline
		Nominal $t\bar{t}$ & -0.2758 & 0.4439 \\
		\hline
		Shifted JES $t\bar{t} $ & -0.1659 & 0.4137  \\
		\hline
	\end{tabular}
	\label{tab:JESimpact}
\end{table} 

\subsubsection{Likelihood scan} 
A likelihood can be built from the MEM probabilities as
\begin{eqnarray}
L(x|\alpha) &=& \prod_{i=1}^{n} P(x_i|\alpha),
\label{eqn:likelihood} 
\end{eqnarray}
where the product is over $n$ measured events. This likelihood will peak around the parameter $\alpha$ which can be any measurable physics quantity such as a mass or a coupling and in the context of the $H \rightarrow ZA$ hypothesis will represent ($M_A$,$M_H$). It is expected that events generated from this process will produce a likelihood peaked in the two dimensional parameter space with a width roughly equal to the experimental resolution of the invariant masses $m_{jj}$ and $m_{lljj}$ used as estimators of the parameters if $n=1$ and will improve when more events are taken into account. 

The log-likelihood on simulated events defined as
\begin{eqnarray}
-\log(L(x|\alpha) ) &=& \frac{1}{n}\sum_{i=1}^{n} -\log(W(x_i|\alpha)) +  \log(\sigma_{\alpha}^{vis}),
\label{eqn:loglikelihood} 
\end{eqnarray}
where the geometric mean of the likelihood is used to evaluate the resolution that would be obtained for one measured $H \rightarrow ZA$ event with no background. In that expression, $-\log(W(x_i|\alpha))$ is the output of the DNN. Had it been computed with MoMEMta, each event would have had to be integrated for several values of $\alpha = (M_A,M_H)$ with a granularity fine enough to allow the fit of a potential peak. In this particular case the computation time will grow exponentially with the parameter space dimentionality. In our two-dimensional case it is already a major pitfall. 

The DNN on the other hand can evaluate any event unseen during training on a grid of parameter points with the same event inputs. The grid can be made arbitrarily fine due to the non-linear interpolation property of the DNN for a very low cost in computation time. The number of evaluations still scales exponentially with the parameter space dimensionality but evaluating the DNN on batches of event allows to take advantage of parallelization to break the exponential dependence.

The second term of Equation~\ref{eqn:loglikelihood} is important and must be evaluated separately. In our specific case the definition given at Section~\ref{sec:MEM} translates to 
\begin{eqnarray}
\sigma_{\alpha}^{vis} = \textrm{Acceptance} \times \sigma(H \ \textrm{production}) &\times& BR(H \rightarrow ZA) \times BR(A \rightarrow bb) \times BR(Z \rightarrow ll).
\label{eqn:visible_xsec}
\end{eqnarray}

\begin{figure}[htp]
	\centering
	\includegraphics[width=.5\textwidth]{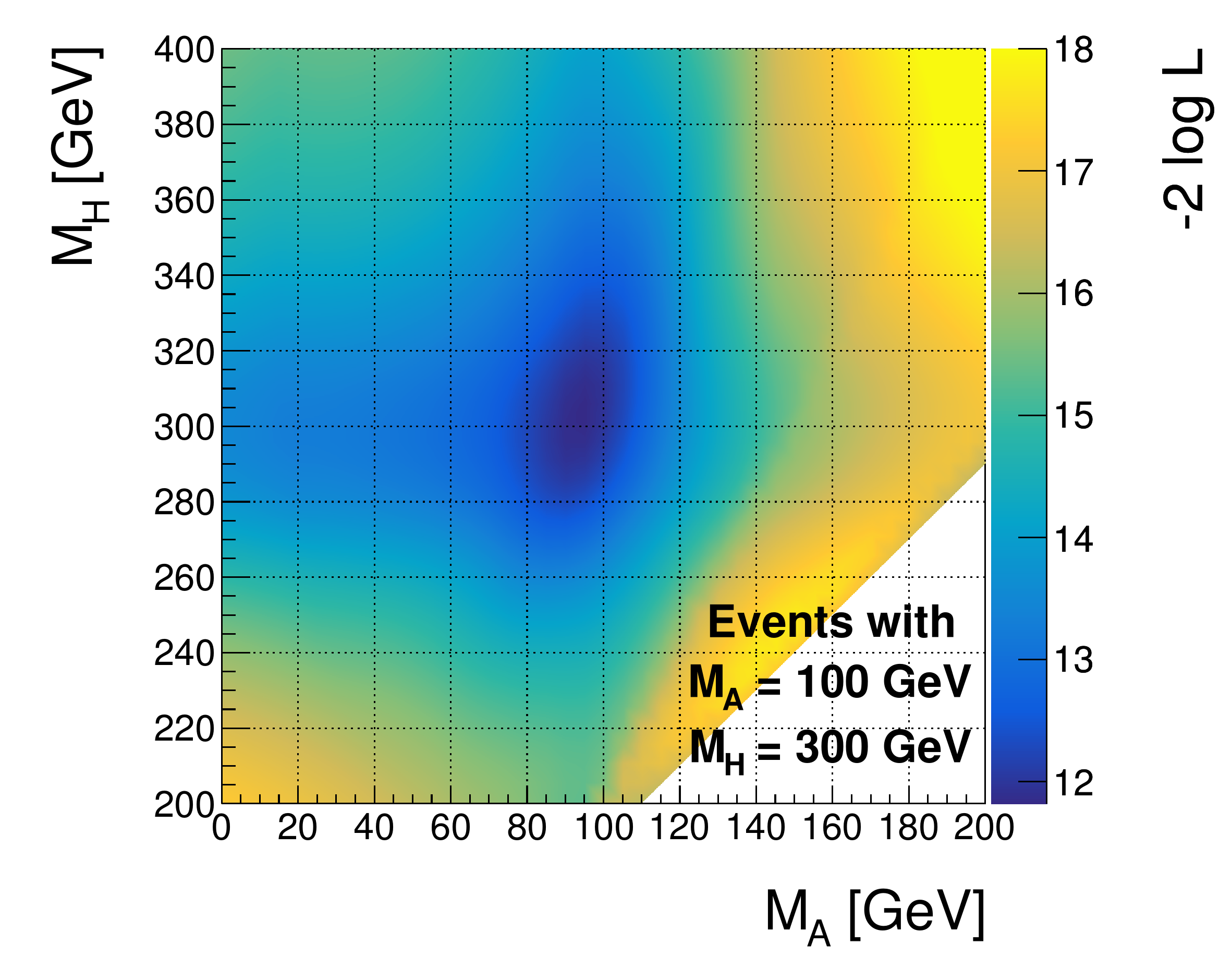}\hfill
	\includegraphics[width=.5\textwidth]{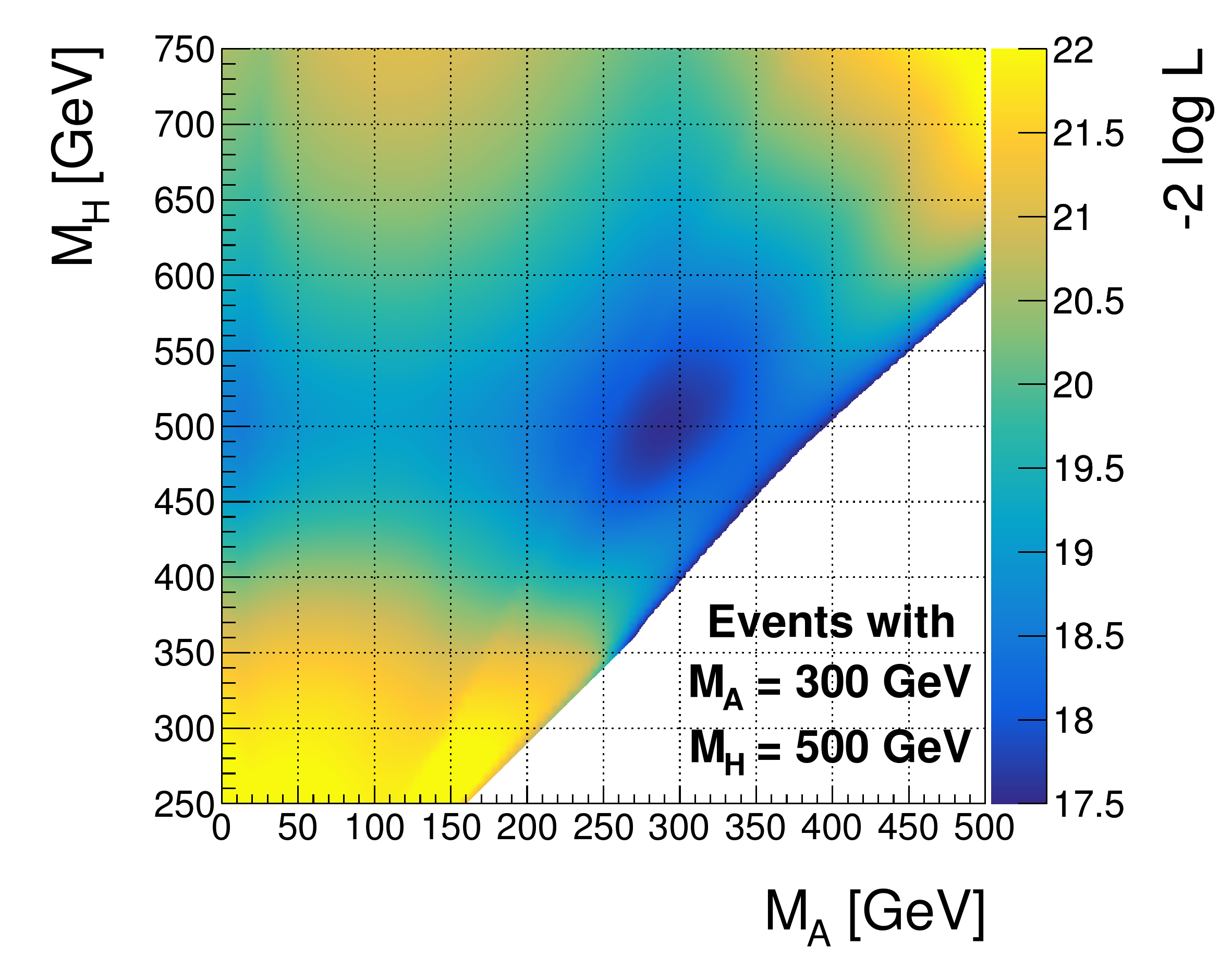}
	\caption{Log-likelihood scan from two different $H \rightarrow ZA$ samples}
	\label{fig:likelihood}
\end{figure}

The acceptance is measured on simulation and the theoretical cross-section must be multiplied by the branching ratios of the particular channel being looked at. The production cross section and branching ratio $A \rightarrow bb$ are very model dependent and are not taken into account. The branching ratio of $H \rightarrow ZA$ is mostly kinematic dependent and was kept, however a model dependent effect can be seen when $M_H = 2 \times M_A$, when the $H \rightarrow AA$ process becomes relevant.
The resulting likelihood is presented in Figure~\ref{fig:likelihood}.

\begin{figure}[htp]
	\centering
	\includegraphics[width=\textwidth]{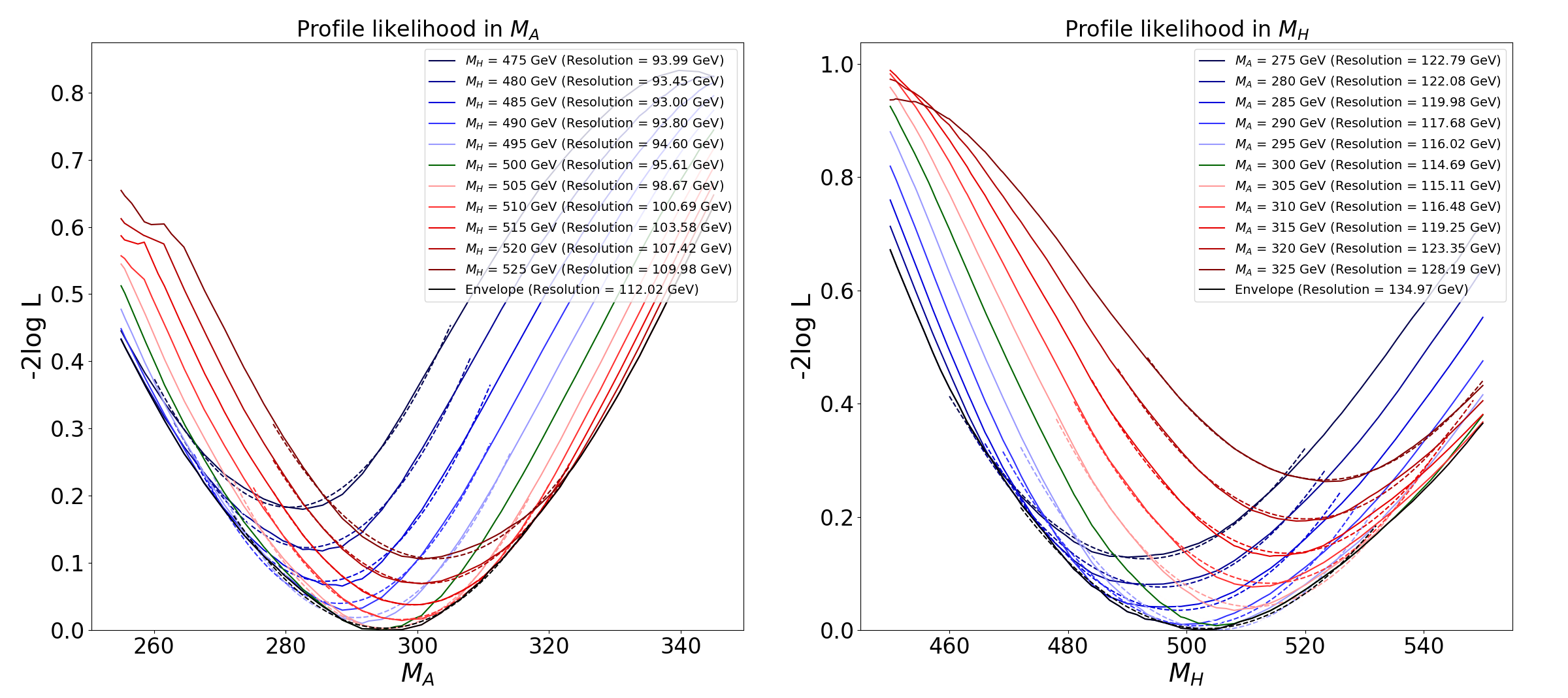} 
	\caption{Profile likelihoods averaged over the number of events (minimum scaled to zero for easier interpretation) and  produced with a mass configuration of $m_H =\SI{500}{\GeV}$ and $m_A = \SI{300}{\GeV}$. Each profile has been computed for several values of the other parameter, the green curve is the one obtained with the parameter used in the event production and the blue and red curves are for smaller and larger (respectively) values. Each solid line is accompanied by a shorter dotted line illustrating the polynomial fit applied on this portion of data which is used to compute the 1-sigma resolution in the legend.}
	\label{fig:resolution}
\end{figure}

The profile likelihoods in both $M_H$ and $M_A$ are shown in Figure~\ref{fig:resolution}. For each profile, several values of the other parameter are displayed to detect an eventual shift in the expected peak. The central part of each curve is used to fit a second order polynom to obtain the resolution of the likelihood peak and compare it to the mass resolution. 
This likelihood has been built by averaging the contributions of the simulated $H \rightarrow ZA$ events, emulating what would be observed on a single measured $H \rightarrow ZA$ event. The computed resolutions are compatible with the experimental resolution of the invariant mass computed from the reconstructed leptons and jets ($m_{jj}$ and $m_{lljj}$). 
As opposed to  these estimators of $M_H$ and $M_A$, the likelihood is unbiased and provides a more proper way of studying an observed resonance. 

\subsection{Direct application to a real-life analysis}
\begin{figure}[htp]
	\centering
	\includegraphics[width=0.8\textwidth]{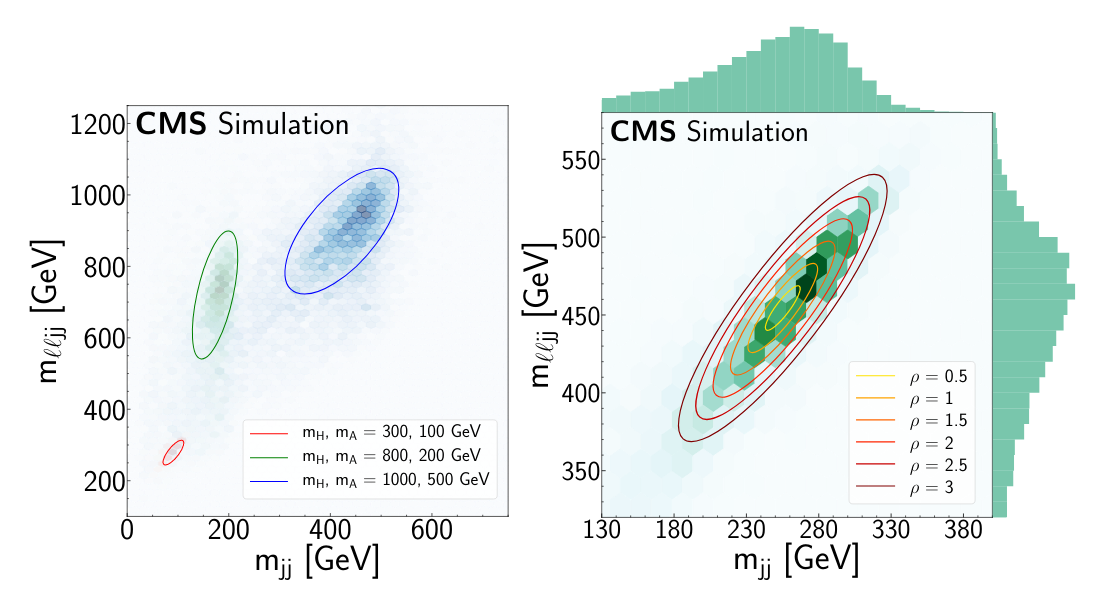} 
	\caption{Signal distribution in the ($m_{jj}$,$m_{lljj}$) plane for several mass configurations (left) and several elliptic fit sizes (parameterized by $\rho$) at a specific mass point of $m_H = \SI{500}{\GeV}$ and $m_A = \SI{300}{\GeV}$ (right), to be compared with the 2D distribution in the plane. From~\cite{HZA}.}
	\label{fig:ellipse_resonance}
\end{figure}

As an example of a real-life study that tackles the $llb\bar{b}$ final states, we will look closer to the CMS $H\rightarrow ZA \rightarrow llb\bar{b}$ analysis~\cite{HZA}. The strategy of this study is to exploit the kinematics of the process $H \rightarrow Z (\rightarrow l^+ l^-) A (\rightarrow b \bar{b})$ (where $l^- = e^- \ or \  \mu^-$) to reconstruct the masses of both H and A bosons using the two- and four-body invariant masses, $m_{jj}$ and $m_{lljj}$ and to define a signal region using these quantities. These distributions are positively correlated and an elliptic signal region has been chosen. The sizes and tilt angles depend on the kinematics and the masses themselves. Hence the parameterization that this analysis used is based on one-dimensional Gaussian fits of the $m_{jj}$ and $m_{lljj}$ distributions to obtain the reconstructed center and the diagonalization of the covariance matrix of both distributions yields the axes and tilt angle (Figure~\ref{fig:ellipse_resonance}). The fraction of signal and background events in different ellipses sizes is binned to be used in a maximum likelihood procedure. 
The use of ellipses in this kind of searches makes it very well optimized and hard to improve without loss of generality.
The MEM is not expected to surpass the standard approach for such an analysis, but should be able to approach the published performances.

Throughout our paper so far, only 23 points were used in the DNN training. As in the CMS paper, our models will be applied on a larger ($>200$) set of $H \rightarrow ZA$ samples to cover finely the mass plane.
Since no retraining takes place, we do not have to worry about overfitting issues.

\begin{figure}[htp]
	\centering
	\includegraphics[width=.5\textwidth]{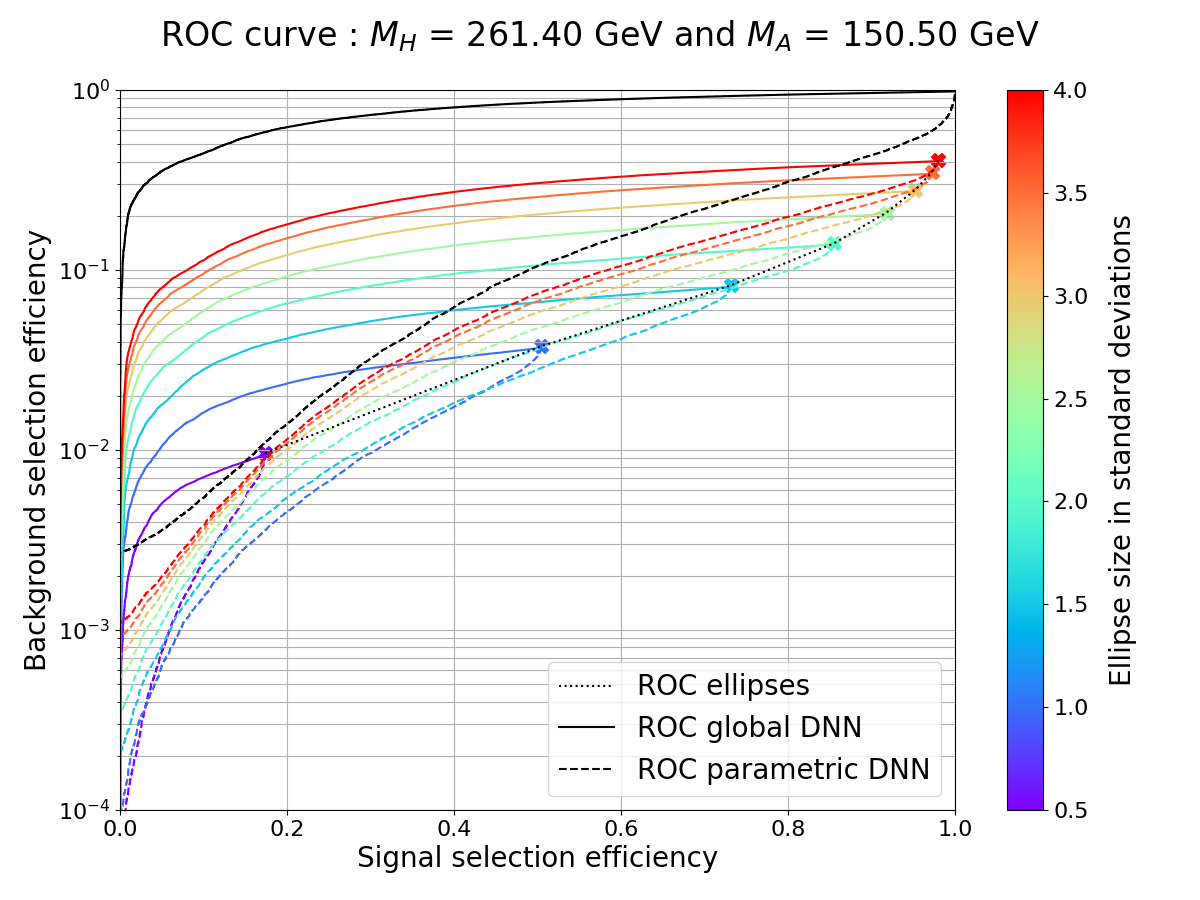}\hfill
	\includegraphics[width=.5\textwidth]{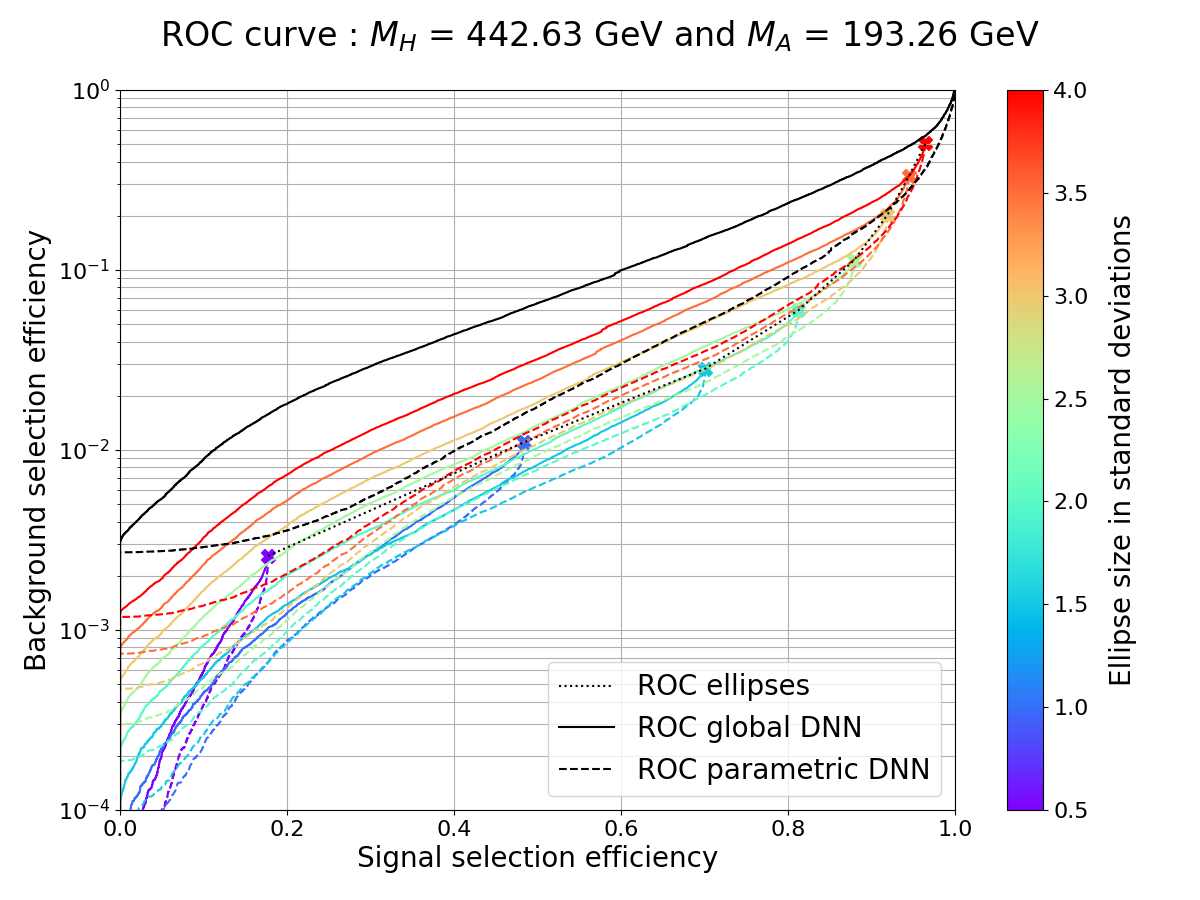}
	\caption{Comparison between the ROC curves of the ellipse method for two mass points with different sizes parameterized by the scale factor $\rho$ (in dotted line), both the global (solid line) and parameterized (dashed line) classifiers, and the combination of these methods. Each ellipse is denoted by a marker and the events that pass its selection are then used to compute the ROC curve with both classifiers.}
	\label{fig:ROC_ellipse}
\end{figure}

Every signal and background event has been processed through the regressive networks of Sections~\ref{sec:DY_weights}, \ref{sec:TT_weights} and~\ref{sec:signal_weights} in order to produce the weights needed by the two classifiers (global and parametric) that were then applied for each mass configuration.
The ROC curve has then been computed for each classifier either using all preselected events, or using only events contained in the elliptic region defined in the mass plane for a given mass hypothesis and a give size parameter  $\rho$.
To do so, the script provided by the collaboration~\cite{HZA} has been used. 
Results are presented in Figure~\ref{fig:ROC_ellipse} for two representative mass points, and compared to the performance of the elliptic selection alone.

\begin{table}[htp]
	\caption{Average computation time in MoMEMta for each type of weights and events. For Drell-Yan and $t\bar{t}$ weights the time distribution is peaked around the values given below, while for the $H \rightarrow ZA$ weights it is much wider - it is very common to find a computation time twice of three times the value given here. The sizes of the samples involved in this section are given as an indication of the overall computation time, the $H \rightarrow ZA$ one includes all 207 masses.}
	\centering
	\renewcommand{\arraystretch}{1.1}
	\begin{tabular}{|c|c|c|c|}
		\hline 
		& Drell-Yan events & $t\bar{t}$ events & $H \rightarrow ZA$ events \\ 
		\hline 
		Drell-Yan weights & \SI{3.6}{\second} / event & \SI{3.8}{\second} / event & \SI{4}{\second} / event \\ 
		\hline 
		$t\bar{t}$ weights &  \SI{12}{\second} / event & \SI{10}{\second}/ event & \SI{20}{\second} / event \\ 
		\hline 
		$H \rightarrow ZA$ weights&  $\sim$ \SI{600}{\second} / event  &$\sim$ \SI{600}{\second} / event &$\sim$ \SI{600}{\second} / event \\  
		\hline 
		Sample size & 279203 events& 461244 events  &2570596  events \\ 
		\hline 
	\end{tabular} 
	\label{tab:time}
\end{table} 

As expected, the global classifier brings no improvement to the ellipse method - especially at low masses. In some cases the combined ROC curve shows a potential gain by taking a larger ellipse combined with the classifier, as can be seen on the right plot. However this improvement is mostly located in the high purity region which is not the one aimed for in the CMS analysis. The ellipse method is very well optimized in the search for a resonance while the global classifier searches for an excess in the whole mass plane. As already mentioned, the latter will require a larger excess to detect something but will not be as heavily affected by the look-elsewhere effect.

While better, the parametric network alone will only outperform the standard approach at very high masses when the ellipses are ill-defined. However, the combination of both becomes interesting for lower masses. In some cases the background contamination can be reduced by almost one order of magnitude with very small loss in signal efficiency. This effect is visible for $M_H$ as low as \SI{200}{\GeV} and increases towards the boosted and high mass regions. 

\begin{table}[htp]
	\caption{Total hypothetical computation time if all the weights had to be computed with MoMEMta and not the DNN in order to be used by the classifiers. For each sample and each weight the number of times MoMEMta would have been called is given as well as the computation time it would have required. The number of calls depends on the type of classifier. The global classifier requires only one weight per SM process but 23 different $H \rightarrow ZA$ weights, while the SM weights need to be evaluated 207 times for the parametric classifier --- because they are repeated at every mass point --- but only once for the $H \rightarrow ZA$ weight. The total time per classifier and process is also given in years.}
	\centering
	\renewcommand{\arraystretch}{1.1}
	\resizebox{\textwidth}{!}{
		\begin{tabular}{|c|c|c|c|c|c|c|}
			\hline 
			Process & \multicolumn{2}{c|}{Drell-Yan events} & \multicolumn{2}{|c|}{ $t\bar{t}$ events} & \multicolumn{2}{c|}{$H \rightarrow ZA$ events} \\ 
			\hline 
			\hline
			\textbf{Global classifier} & Calls (x1) & Time [days] &  Calls (x1) & Time [days]  &  Calls (x1) & Time [days] \\
			\hline 
			Drell-Yan weights (x1) & 279 203 & 11.6 & 461 244 & 20.3 & 2 570 596 & 119.0 \\ 
			\hline 
			$t\bar{t}$ weights (x1) & 279 203 & 32.3 & 461 244 & 85.4 & 2 570 596 & 595.0 \\ 
			\hline 
			$H \rightarrow ZA$ weights (x23) & 6 421 669 & 44 594.9 & 10 608 612 & 73670.9 & 59 123 087 & 410 581.3 \\ 
			\hline
			\hline
			Total time [years] & \multicolumn{2}{c|}{122.3} & \multicolumn{2}{|c|}{202.1} & \multicolumn{2}{c|}{1126.8} \\
			\hline
			\hline
			\textbf{Parametric classifier} & Calls (x207) & Time [days] &  Calls (x207) & Time [days]  &  Calls (x1) & Time [days] \\
			\hline 
			Drell-Yan weights (x1) & 57 795 021 & 2408.1 & 95 477 508 & 4199.2 & 2 570 596 & 119.0 \\ 
			\hline 
			$t\bar{t}$ weights (x1) & 57 795 021 & 8027,1 & 95 477 508 & 17681.0 & 2 570 596  & 595.0 \\ 
			\hline 
			$H \rightarrow ZA$ weights (x1) & 57 795 021  & 401 354.3 & 95 477 508 & 663 038.2 & 2 570 596  & 17 851.4 \\ 
			\hline
			\hline
			Total time [years] & \multicolumn{2}{c|}{1128.2} & \multicolumn{2}{|c|}{1876.5} & \multicolumn{2}{c|}{50.9} \\
			\hline		
	\end{tabular}}
	\label{tab:calls}
	
\end{table}

While the simple approach followed in this work only has the potential to marginally improve the CMS $H \rightarrow ZA$ analysis, the DNN ansatz still opens a wide range of possibilities previously out-of-reach of the MEM. 
Apart from their training time, which have lasted from a few hours to a single day on CPU, evaluating a weight or a probability is a very fast process : about $150 \mu s$ with large enough batches, with small variations depending on the depth of the network. This must be compared to the computation time of MoMEMta (Table~\ref{tab:time}) and the number of times it would have been called to produce Figure~\ref{fig:ROC_ellipse}
 (Table~\ref{tab:calls}). 
The weight computations for the global (parametric) classifier would have taken about 1450 (3050) CPU years, which is more than prohibitive even with a large farm of CPU. Using the DNN to produce the weights requires in total less than 10 hours, where must be added the time of I/O data streams, RAM allocation, data repetition and processing --- all of which even tend to become dominant compared to the pure weight production. 
These weights must be fed to the classifiers, looped through for each ellipse and the ROC curves must be computed. In the end, with the DNN, the production of Figure~\ref{fig:ROC_ellipse} on a cluster of CPUs with a few hundred nodes took less than a day. The pure weight computation has been reduced by six orders of magnitude.

\section{Conclusion}
In this paper we presented a method where the integral of the Matrix Element Method is regressed by a Deep Neural Network in order to speed up the computations involved in the MEM.
From the few representative processes studied in this paper, we conclude that a DNN can be trained that closely reproduces the results of the direct numerical integration of the matrix element using dedicated tools like MoMEMta.
This regression with the DNN introduces inevitable inaccuracies in the weights that nonetheless do not have a significantly impact on the performance in the studied applications.
Faster weight calculations open a wide range of possibilities : study of systematics, likelihood scans, parameters scans and in general enables the use of the MEM for a new wider class of physics analyses, including the search for new physics.

\section{Acknowledgments}
We warmly thank Olivier Mattelaer for his feedback about the method itself, Sebastien Wertz for his help with MoMEMta and pertinent remarks, Alessia Saggio for her many contributions regarding the $H\rightarrow ZA$ analysis and Pieter David for his helpful comments. Florian Bury is a Research Fellow of the Fonds de la Recherche Scientifique – FNRS. Christophe Delaere is Senior Research Associate of the FNRS.
Part of this work has been funded by the IISN under convention 4.4503.16.
Computational resources have been provided by the supercomputing facilities of the Université catholique de Louvain (CISM/UCL) and the Consortium des Équipements de Calcul Intensif en Fédération Wallonie Bruxelles (CÉCI) funded by the Fond de la Recherche Scientifique de Belgique (F.R.S.-FNRS) under convention 2.5020.11.

\clearpage
\bibliographystyle{utphys}  
\bibliography{references} 

\providecommand{\href}[2]{#2}\begingroup\raggedright\begin{thebibliography}{10}

\bibitem{pivot}
G.~{Louppe}, M.~{Kagan}, and K.~{Cranmer}, ``{Learning to Pivot with
  Adversarial Networks},'' {\em arXiv e-prints} (Nov., 2016) arXiv:1611.01046,
  \href{http://arxiv.org/abs/1611.01046}{{\ttfamily arXiv:1611.01046
  [stat.ML]}}.

\bibitem{WeakSupervised}
L.~M. {Dery}, B.~{Nachman}, F.~{Rubbo}, and A.~{Schwartzman}, ``{Weakly
  supervised classification in high energy physics},''
  \href{http://dx.doi.org/10.1007/JHEP05(2017)145}{{\em Journal of High Energy
  Physics} {\bfseries 2017} no.~5, (May, 2017) 145},
  \href{http://arxiv.org/abs/1702.00414}{{\ttfamily arXiv:1702.00414
  [hep-ph]}}.

\bibitem{SemiSupervised}
M.~{Kuusela}, T.~{Vatanen}, E.~{Malmi}, T.~{Raiko}, T.~{Aaltonen}, and
  Y.~{Nagai},
  \href{http://dx.doi.org/10.1088/1742-6596/368/1/012032}{``{Semi-supervised
  anomaly detection - towards model-independent searches of new physics},''} in
  {\em Journal of Physics Conference Series}, vol.~368 of {\em Journal of
  Physics Conference Series}, p.~012032.
\newblock June, 2012.
\newblock \href{http://arxiv.org/abs/1112.3329}{{\ttfamily arXiv:1112.3329
  [physics.data-an]}}.

\bibitem{Unsupervised}
A.~{Andreassen}, I.~{Feige}, C.~{Frye}, and M.~D. {Schwartz}, ``{JUNIPR: a
  framework for unsupervised machine learning in particle physics},''
  \href{http://dx.doi.org/10.1140/epjc/s10052-019-6607-9}{{\em European
  Physical Journal C} {\bfseries 79} no.~2, (Feb., 2019) 102},
  \href{http://arxiv.org/abs/1804.09720}{{\ttfamily arXiv:1804.09720
  [hep-ph]}}.

\bibitem{MEM1}
R.~Dalitz and G.~R. Goldstein, ``{Test of analysis method for top-antitop
  production and decay events},''
  \href{http://dx.doi.org/10.1098/rspa.1999.0428}{{\em Proc. Roy. Soc. Lond. A}
  {\bfseries A455} (1999) 2803--2834},
  \href{http://arxiv.org/abs/hep-ph/9802249}{{\ttfamily arXiv:hep-ph/9802249}}.

\bibitem{MEM2}
{CDF Collaboration}, ``{Top Quark Mass Measurement in the $t \bar{t}$ All
  Hadronic Channel using a Matrix Element Technique in $p\bar{p}$ Collisions at
  $\sqrt{s}$ = 1.96-TeV},''
  \href{http://dx.doi.org/10.1103/PhysRevD.79.072010}{{\em Phys. Rev. D}
  {\bfseries 79} (2009) 072010},
  \href{http://arxiv.org/abs/0811.1062}{{\ttfamily arXiv:0811.1062 [hep-ex]}}.

\bibitem{MEM3}
{CDF Collaboration}, ``{Measurement of the top quark mass with dilepton events
  selected using neuroevolution at CDF},''
  \href{http://dx.doi.org/10.1103/PhysRevLett.102.152001}{{\em Phys. Rev.
  Lett.} {\bfseries 102} (2009) 152001},
  \href{http://arxiv.org/abs/0807.4652}{{\ttfamily arXiv:0807.4652 [hep-ex]}}.

\bibitem{MEM4}
{CDF Collaboration}, ``{Measurement of the top-quark mass in the lepton+jets
  channel using a matrix element technique with the CDF II detector},''
  \href{http://dx.doi.org/10.1103/PhysRevD.84.071105}{{\em Phys. Rev. D}
  {\bfseries 84} (2011) 071105},
  \href{http://arxiv.org/abs/1108.1601}{{\ttfamily arXiv:1108.1601 [hep-ex]}}.

\bibitem{MEM5}
{CDF Collaboration}, ``{Measurements of the Top-quark Mass and the $t\bar{t}$
  Cross Section in the Hadronic $\tau +$ Jets Decay Channel at $\sqrt{s} =
  1.96$ TeV},'' \href{http://dx.doi.org/10.1103/PhysRevLett.109.192001}{{\em
  Phys. Rev. Lett.} {\bfseries 109} (2012) 192001},
  \href{http://arxiv.org/abs/1208.5720}{{\ttfamily arXiv:1208.5720 [hep-ex]}}.

\bibitem{MEM6}
{D0 Collaboration}, ``Precision measurement of the top-quark mass
  inlepton+jetsfinal states,''
  \href{http://dx.doi.org/10.1103/physrevd.91.112003}{{\em Physical Review D}
  {\bfseries 91} no.~11, (Jun, 2015) },
  \href{http://arxiv.org/abs/1501.07912}{{\ttfamily 1501.07912}}.
  \url{http://dx.doi.org/10.1103/PhysRevD.91.112003}.

\bibitem{MEM7}
{D0 Collaboration}, ``Measurement of the top quark mass using the matrix
  element technique in dilepton final states,''
  \href{http://dx.doi.org/10.1103/physrevd.94.032004}{{\em Physical Review D}
  {\bfseries 94} no.~3, (Aug, 2016) },
  \href{http://arxiv.org/abs/1606.02814}{{\ttfamily 1606.02814}}.
  \url{http://dx.doi.org/10.1103/PhysRevD.94.032004}.

\bibitem{MEM8}
{ATLAS Collaboration}, ``{Search for the Standard Model Higgs boson produced in
  association with top quarks and decaying into $b\bar{b}$ in pp collisions at
  $\sqrt{s}$ = 8 TeV with the ATLAS detector},''
  \href{http://dx.doi.org/10.1140/epjc/s10052-015-3543-1}{{\em Eur. Phys. J. C}
  {\bfseries 75} no.~7, (2015) 349},
  \href{http://arxiv.org/abs/1503.05066}{{\ttfamily arXiv:1503.05066
  [hep-ex]}}.

\bibitem{MEM9}
{ATLAS Collaboration}, ``{Search for the standard model Higgs boson produced in
  association with top quarks and decaying into a $b\bar{b}$ pair in $pp$
  collisions at $\sqrt{s}$ = 13 TeV with the ATLAS detector},''
  \href{http://dx.doi.org/10.1103/PhysRevD.97.072016}{{\em Phys. Rev. D}
  {\bfseries 97} no.~7, (2018) 072016},
  \href{http://arxiv.org/abs/1712.08895}{{\ttfamily arXiv:1712.08895
  [hep-ex]}}.

\bibitem{MEM10}
{CMS Collaboration}, ``{Search for $ \mathrm{t}\overline{\mathrm{t}}\mathrm{H}
  $ production in the $ \mathrm{H}\to \mathrm{b}\overline{\mathrm{b}} $ decay
  channel with leptonic $ \mathrm{t}\overline{\mathrm{t}} $ decays in
  proton-proton collisions at $ \sqrt{s}=13 $ TeV},''
  \href{http://dx.doi.org/10.1007/JHEP03(2019)026}{{\em JHEP} {\bfseries 03}
  (2019) 026}, \href{http://arxiv.org/abs/1804.03682}{{\ttfamily
  arXiv:1804.03682 [hep-ex]}}.

\bibitem{MEM12}
{CMS Collaboration}, ``{Search for $\mathrm{t}\overline{\mathrm{t}}$H
  production in the all-jet final state in proton-proton collisions at
  $\sqrt{s}=$ 13 TeV},'' \href{http://dx.doi.org/10.1007/JHEP06(2018)101}{{\em
  JHEP} {\bfseries 06} (2018) 101},
  \href{http://arxiv.org/abs/1803.06986}{{\ttfamily arXiv:1803.06986
  [hep-ex]}}.

\bibitem{MEM13}
{ATLAS Collaboration}, ``{Measurement of the Higgs boson coupling properties in
  the $H\rightarrow ZZ^{*} \rightarrow 4\ell$ decay channel at $\sqrt{s}$ = 13
  TeV with the ATLAS detector},''
  \href{http://dx.doi.org/10.1007/JHEP03(2018)095}{{\em JHEP} {\bfseries 03}
  (2018) 095}, \href{http://arxiv.org/abs/1712.02304}{{\ttfamily
  arXiv:1712.02304 [hep-ex]}}.

\bibitem{MEM14}
{ATLAS Collaboration}, ``{Evidence for the associated production of the Higgs
  boson and a top quark pair with the ATLAS detector},''
  \href{http://dx.doi.org/10.1103/PhysRevD.97.072003}{{\em Phys. Rev. D}
  {\bfseries 97} no.~7, (2018) 072003},
  \href{http://arxiv.org/abs/1712.08891}{{\ttfamily arXiv:1712.08891
  [hep-ex]}}.

\bibitem{MEM15}
{CMS Collaboration}, ``{Evidence for associated production of a Higgs boson
  with a top quark pair in final states with electrons, muons, and hadronically
  decaying $\tau$ leptons at $\sqrt{s} =$ 13 TeV},''
  \href{http://dx.doi.org/10.1007/JHEP08(2018)066}{{\em JHEP} {\bfseries 08}
  (2018) 066}, \href{http://arxiv.org/abs/1803.05485}{{\ttfamily
  arXiv:1803.05485 [hep-ex]}}.

\bibitem{MEM16}
{ATLAS Collaboration}, ``{Evidence for single top-quark production in the
  $s$-channel in proton-proton collisions at $\sqrt{s}=$8 TeV with the ATLAS
  detector using the Matrix Element Method},''
  \href{http://dx.doi.org/10.1016/j.physletb.2016.03.017}{{\em Phys. Lett. B}
  {\bfseries 756} (2016) 228--246},
  \href{http://arxiv.org/abs/1511.05980}{{\ttfamily arXiv:1511.05980
  [hep-ex]}}.

\bibitem{MEM17}
{CMS Collaboration}, ``{Measurement of Spin Correlations in $t\bar{t}$
  Production using the Matrix Element Method in the Muon+Jets Final State in
  $pp$ Collisions at $\sqrt{s} =$ 8 TeV},''
  \href{http://dx.doi.org/10.1016/j.physletb.2016.05.005}{{\em Phys. Lett. B}
  {\bfseries 758} (2016) 321--346},
  \href{http://arxiv.org/abs/1511.06170}{{\ttfamily arXiv:1511.06170
  [hep-ex]}}.

\bibitem{MoMEMta}
S.~Brochet, C.~Delaere, B.~Fran{\c c}ois, V.~Lema\^{\i}tre, A.~Mertens,
  A.~Saggio, M.~{Vidal Marono}, and S.~Wertz, ``{MoMEMta, a modular toolkit for
  the Matrix Element Method at the LHC},''
  \href{http://dx.doi.org/10.1140/epjc/s10052-019-6635-5}{{\em Eur.\ Phys.\ J.\
  C} {\bfseries 79} no.~2, (2019) 126},
  \href{http://arxiv.org/abs/1805.08555}{{\ttfamily arXiv:1805.08555
  [hep-ph]}}.

\bibitem{MEMParallel_1}
D.~Schouten, A.~DeAbreu, and B.~Stelzer, ``Accelerated matrix element method
  with parallel computing,''
  \href{http://dx.doi.org/10.1016/j.cpc.2015.02.020}{{\em Computer Physics
  Communications} {\bfseries 192} (2015) 54--59}.
  \url{http://www.sciencedirect.com/science/article/pii/S0010465515000776}.

\bibitem{MEMParallel_2}
G.~Grasseau, D.~Chamont, F.~Beaudette, L.~Bianchini, O.~Davignon,
  L.~Mastrolorenzo, C.~Ochando, P.~Paganini, and T.~Strebler, ``Matrix element
  method for high performance computing platforms,''
  \href{http://dx.doi.org/10.1088/1742-6596/664/9/092009}{{\em Journal of
  Physics: Conference Series} {\bfseries 664} no.~9, (Dec, 2015) 092009}.
  \url{https://doi.org/10.1088/1742-6596/664/9/092009}.

\bibitem{MEMGPU_1}
G.~Grasseau, F.~Beaudette, C.~Perez, A.~Zabi, A.~Chiron, T.~Strebler, and
  G.~Hautreux, ``Deployment of a matrix element method code for the tth channel
  analysis on gpu{\rq}s platform,''
  \href{http://dx.doi.org/10.1051/epjconf/201921406028}{{\em EPJ Web of
  Conferences} {\bfseries 214} (01, 2019) 06028}.

\bibitem{MEMGPU_2}
D.~Schouten, A.~DeAbreu, and B.~Stelzer,
  \href{http://dx.doi.org/10.3204/DESY-PROC-2014-05/20}{``{GPUs for Higgs boson
  data analysis at the LHC using the Matrix Element Method},''} in {\em {GPU
  Computing in High-Energy Physics}}, pp.~109--118.
\newblock 2015.

\bibitem{BDTIntegration}
J.~{Bendavid}, ``{Efficient Monte Carlo Integration Using Boosted Decision
  Trees and Generative Deep Neural Networks},'' {\em arXiv e-prints} (June,
  2017) arXiv:1707.00028, \href{http://arxiv.org/abs/1707.00028}{{\ttfamily
  arXiv:1707.00028 [hep-ph]}}.

\bibitem{DNNIntegration}
M.~D. {Klimek} and M.~{Perelstein}, ``{Neural Network-Based Approach to Phase
  Space Integration},'' {\em arXiv e-prints} (Oct., 2018) arXiv:1810.11509,
  \href{http://arxiv.org/abs/1810.11509}{{\ttfamily arXiv:1810.11509
  [hep-ph]}}.

\bibitem{NormFlow1}
E.~{Bothmann}, T.~{Jan{\ss}en}, M.~{Knobbe}, T.~{Schmale}, and S.~{Schumann},
  ``{Exploring phase space with Neural Importance Sampling},''
  \href{http://dx.doi.org/10.21468/SciPostPhys.8.4.069}{{\em SciPost Physics}
  {\bfseries 8} no.~4, (Apr., 2020) 069},
  \href{http://arxiv.org/abs/2001.05478}{{\ttfamily arXiv:2001.05478
  [hep-ph]}}.

\bibitem{NormFlow2}
C.~{Gao}, J.~{Isaacson}, and C.~{Krause}, ``{i-flow: High-dimensional
  Integration and Sampling with Normalizing Flows},'' {\em arXiv e-prints}
  (Jan., 2020) arXiv:2001.05486,
  \href{http://arxiv.org/abs/2001.05486}{{\ttfamily arXiv:2001.05486
  [physics.comp-ph]}}.

\bibitem{LikelihoodFree_1}
K.~{Cranmer}, J.~{Pavez}, and G.~{Louppe}, ``{Approximating Likelihood Ratios
  with Calibrated Discriminative Classifiers},'' {\em arXiv e-prints} (June,
  2015) arXiv:1506.02169, \href{http://arxiv.org/abs/1506.02169}{{\ttfamily
  arXiv:1506.02169 [stat.AP]}}.

\bibitem{LikelihoodFree_2}
J.~{Brehmer}, G.~{Louppe}, J.~{Pavez}, and K.~{Cranmer}, ``{Mining gold from
  implicit models to improve likelihood-free inference},'' {\em arXiv e-prints}
  (May, 2018) arXiv:1805.12244,
  \href{http://arxiv.org/abs/1805.12244}{{\ttfamily arXiv:1805.12244
  [stat.ML]}}.

\bibitem{LikelihoodFree_3}
M.~{Stoye}, J.~{Brehmer}, G.~{Louppe}, J.~{Pavez}, and K.~{Cranmer},
  ``{Likelihood-free inference with an improved cross-entropy estimator},''
  {\em arXiv e-prints} (Aug., 2018) arXiv:1808.00973,
  \href{http://arxiv.org/abs/1808.00973}{{\ttfamily arXiv:1808.00973
  [stat.ML]}}.

\bibitem{MG5}
J.~Alwall, R.~Frederix, S.~Frixione, V.~Hirschi, F.~Maltoni, O.~Mattelaer,
  H.~S. Shao, T.~Stelzer, P.~Torrielli, and M.~Zaro, ``{The automated
  computation of tree-level and next-to-leading order differential cross
  sections, and their matching to parton shower simulations},''
  \href{http://dx.doi.org/10.1007/JHEP07(2014)079}{{\em JHEP} {\bfseries 07}
  (2014) 079}, \href{http://arxiv.org/abs/1405.0301}{{\ttfamily arXiv:1405.0301
  [hep-ph]}}.

\bibitem{VEGAS}
G.~P. Lepage, ``A new algorithm for adaptive multidimensional integration,''
  \href{http://dx.doi.org/10.1016/0021-9991(78)90004-9}{{\em Journal of
  Computational Physics} {\bfseries 27} no.~2, (1978) 192--203}.

\bibitem{DataPreProcess}
B.~{Hashemi}, N.~{Amin}, K.~{Datta}, D.~{Olivito}, and M.~{Pierini}, ``{LHC
  analysis-specific datasets with Generative Adversarial Networks},'' {\em
  arXiv e-prints} (Jan., 2019) arXiv:1901.05282,
  \href{http://arxiv.org/abs/1901.05282}{{\ttfamily arXiv:1901.05282
  [hep-ex]}}.

\bibitem{outputPreprocess}
A.~{Coccaro}, M.~{Pierini}, L.~{Silvestrini}, and R.~{Torre}, ``{The
  DNNLikelihood: enhancing likelihood distribution with Deep Learning},''
  \href{http://dx.doi.org/10.1140/epjc/s10052-020-8230-1}{{\em European
  Physical Journal C} {\bfseries 80} no.~7, (July, 2020) 664},
  \href{http://arxiv.org/abs/1911.03305}{{\ttfamily arXiv:1911.03305
  [hep-ph]}}.

\bibitem{keras}
F.~Chollet {\em et~al.}, ``Keras.'' \url{https://keras.io}, 2015.

\bibitem{tensorflow}
M.~A. et~al, ``{TensorFlow}: Large-scale machine learning on heterogeneous
  systems,'' 2015.
\newblock \url{http://tensorflow.org/}. Software available from tensorflow.org.

\bibitem{AtlasHZA}
{ATLAS Collaboration}, ``{Search for a heavy Higgs boson decaying into a $Z$
  boson and another heavy Higgs boson in the $\ell\ell bb$ final state in $pp$
  collisions at $\sqrt{s}=13$ TeV with the ATLAS detector},''
  \href{http://dx.doi.org/10.1016/j.physletb.2018.07.006}{{\em Phys. Lett. B}
  {\bfseries 783} (2018) 392--414},
  \href{http://arxiv.org/abs/1804.01126}{{\ttfamily arXiv:1804.01126
  [hep-ex]}}.

\bibitem{HZA}
{CMS Collaboration}, ``{Search for new neutral Higgs bosons through the H$\to$
  ZA $\to \ell^{+}\ell^{-} \mathrm{b\bar{b}}$ process in pp collisions at
  $\sqrt{s} =$ 13 TeV},'' \href{http://dx.doi.org/10.1007/JHEP03(2020)055}{{\em
  JHEP} {\bfseries 03} (2020) 055},
  \href{http://arxiv.org/abs/1911.03781}{{\ttfamily arXiv:1911.03781
  [hep-ex]}}. data on \href{https://doi.org/10.17182/hepdata.90710}{HEPData}.

\bibitem{paramDNN}
P.~Baldi, K.~Cranmer, T.~Faucett, P.~Sadowski, and D.~Whiteson,
  ``{Parameterized neural networks for high-energy physics},''
  \href{http://dx.doi.org/10.1140/epjc/s10052-016-4099-4}{{\em Eur. Phys. J. C}
  {\bfseries 76} no.~5, (2016) 235},
  \href{http://arxiv.org/abs/1601.07913}{{\ttfamily arXiv:1601.07913
  [hep-ex]}}.

\bibitem{relu}
A.~F. Agarap, ``Deep learning using rectified linear units (relu),'' {\em CoRR}
  {\bfseries abs/1803.08375} (2018) ,
  \href{http://arxiv.org/abs/1803.08375}{{\ttfamily arXiv:1803.08375}}.
  \url{http://arxiv.org/abs/1803.08375}.

\bibitem{selu}
D.~Pedamonti, ``Comparison of non-linear activation functions for deep neural
  networks on {MNIST} classification task,'' {\em CoRR} {\bfseries
  abs/1804.02763} (2018) , \href{http://arxiv.org/abs/1804.02763}{{\ttfamily
  arXiv:1804.02763}}. \url{http://arxiv.org/abs/1804.02763}.

\bibitem{adam}
D.~P. {Kingma} and J.~{Ba}, ``{Adam: A Method for Stochastic Optimization},''
  {\em arXiv e-prints} (Dec., 2014) arXiv:1412.6980,
  \href{http://arxiv.org/abs/1412.6980}{{\ttfamily arXiv:1412.6980 [cs.LG]}}.

\bibitem{dropout}
S.~Nitish, G.~Hinton, A.~Krizhevsky, I.~Sutskever, and R.~Salakhutdinov,
  ``Dropout: A simple way to prevent neural networks from overfitting,'' {\em
  Journal of Machine Learning Research} {\bfseries 15} (2014) 1929--1958.

\bibitem{l2}
C.~Cortes, M.~Mohri, and A.~Rostamizadeh, ``{L2} regularization for learning
  kernels,'' {\em CoRR} {\bfseries abs/1205.2653} (2012) ,
  \href{http://arxiv.org/abs/1205.2653}{{\ttfamily arXiv:1205.2653}}.
  \url{http://arxiv.org/abs/1205.2653}.

\end{thebibliography}\endgroup

\end{document}